\pdfoutput=1

\documentclass[11pt]{article}
\PassOptionsToPackage{table}{xcolor} 

\usepackage[final]{acl}
\usepackage{float}

\usepackage{times}
\usepackage{latexsym}
\usepackage{pifont}
\usepackage{multirow}

\usepackage[normalem]{ulem}
\useunder{\uline}{\ul}{}
\definecolor{orange}{RGB}{243, 204, 174}

\usepackage{subcaption}
\usepackage[T1]{fontenc}

\usepackage[utf8]{inputenc}
\usepackage{url} 
\usepackage{amsthm}

\usepackage{microtype}
\usepackage{xcolor}
\usepackage[table]{xcolor}
\definecolor{green}{RGB}{199, 223, 181}
\definecolor{orange}{RGB}{243, 204, 174}
\definecolor{blue1}{RGB}{193, 214, 237}
\definecolor{orange4table}{RGB}{243, 204, 174}
\definecolor{lightpurple}{RGB}{232, 229, 250}
\usepackage{multirow}
\usepackage{booktabs}

\usepackage{inconsolata}

\usepackage{hyperref}

\theoremstyle{plain}
\newtheorem{theorem}{Theorem}[section]

\theoremstyle{definition}

\newtheorem{assumption}[theorem]{Assumption}
\theoremstyle{remark}

\usepackage{graphicx}
\usepackage{amsmath}
\usepackage{amssymb}
\usepackage{mathtools}
\usepackage{amsthm}
\usepackage{booktabs}
\usepackage{tcolorbox}
\tcbset{
  colframe=teal,       
  colback=green!20,      
  boxrule=0.3mm,        
  arc=3mm,              
  left=3mm,             
  right=3mm,            
  top=2mm,              
  bottom=2mm,           
}
\usepackage[capitalize,noabbrev]{cleveref}

\usepackage{ifthen}
\usepackage[normalem]{ulem} 
\usepackage{xcolor}

\newboolean{showedits}
\setboolean{showedits}{true} 
\ifthenelse{\boolean{showedits}}
{
	\newcommand{\del}[1]{\textcolor{red}{\sout{#1}}} 
}{
	\newcommand{\del}[1]{} 
	
}

\newboolean{showcomments}
\setboolean{showcomments}{true} 
\newcommand{\id}[1]{$-$Id: scgPaper.tex 32478 2010-04-29 09:11:32Z oscar $-$}

\ifthenelse{\boolean{showcomments}}
{\newcommand{\nbc}[3]{
		{\colorbox{#3}{\bfseries\sffamily\scriptsize\textcolor{white}{#1}}}
		{\textcolor{#3}{$\blacktriangleright$#2$\blacktriangleleft$}}}
	}
{\newcommand{\nbc}[3]{}
	\renewcommand{\del}[1]{} 
	}

\definecolor{ibcolor}{rgb}{1.0,0.2,.4}
\definecolor{dsrcolor}{rgb}{0.5,0.6,0}
\definecolor{cfcolor}{rgb}{0,0.5,0.9}
\definecolor{oldcolor}{rgb}{0.2,0.2,0.2}
\definecolor{tdcolor}{rgb}{1.0,0,0}
\definecolor{oldcolor}{rgb}{0.5,0.5,0.5}
\definecolor{lycolor}{rgb}{0.3,0.3,0.8}

\title{Multi-task Adversarial Attacks against Black-box Model \\with Few-shot Queries}

\author{
\textbf{Wenqiang Wang$^{1}$, Yan Xiao$^{1}$, Hao Lin$^{1}$,  Yangshijie Zhang$^{3}$,  Xiaochun Cao$^{1,2}$\thanks{Corresponding author} ,} \\
  $^{1}$Shenzhen Campus of Sun Yat-sen University, $^{2}$ Peng Cheng Laboratory
$^{3}$Lanzhou University \\
\texttt{wangwq69@mail2.sysu.edu.cn,~xiaoy367@mail.sysu.edu.cn,}\\ 
\texttt{~linh255@mail2.sysu.edu.cn,~zhangyshj2023@lzu.edu.cn,}\\
\texttt{~caoxiaochun@mail.sysu.edu.cn}
}

\begin{document}
\maketitle
\begin{abstract}
Current multi-task adversarial text attacks rely on abundant access to shared internal features and numerous queries, often limited to a single task type. As a result, these attacks are less effective against practical scenarios involving black-box feedback APIs, limited queries, or multiple task types.
To bridge this gap, we propose 
\textbf{C}luster and \textbf{E}nsemble \textbf{M}ulti-task Text Adversarial \textbf{A}ttack (\textbf{CEMA}), an effective black-box attack that exploits the transferability of adversarial texts across different tasks.
CEMA simplifies complex multi-task scenarios by using a \textit{deep-level substitute model} trained in a \textit{plug-and-play} manner for text classification, enabling attacks without mimicking the victim model. This approach requires only a few queries for training, converting multi-task attacks into classification attacks and allowing attacks across various tasks. 

CEMA generates multiple adversarial candidates using different text classification methods and selects the one that most effectively attacks substitute models. 

In experiments involving multi-task models with two, three, or six tasks—spanning classification, translation, summarization, and text-to-image generation—CEMA demonstrates significant attack success with as few as 100 queries.
Furthermore, CEMA can target commercial APIs (e.g., Baidu and Google Translate), large language models (e.g., ChatGPT 4o), and image-generation models (e.g., Stable Diffusion V2), showcasing its versatility and effectiveness in real-world applications.

\end{abstract}

\section{Introduction}
A multi-task textual adversarial attack targets multiple tasks with minimal perturbations, posing significant risks to safety-critical systems by inducing erroneous decisions~\citep{liu2017adversarial,lin2022multi,watkins2024adversarially}.
Most research on multi-task adversarial examples focuses on tasks of the same type, particularly classification tasks~\citep{liu2017adversarial}. Additionally, existing adversarial attack methods generally assume that attackers have access to the model architecture and shared layer information within a unified model, requiring abundant queries to the victim model~\citep{guo2020multi,zhe2024adversarial}. 
However, these studies do not align well with current real-world scenarios. For instance, consider the prompt ``Translate this paper into Chinese and extract its core meaning:'' input into ChatGPT-4o. This creates a multi-task model with distinct tasks—translation and summarization—where the AI agent (e.g., ChatGPT-4o) is a black-box victim model that outputs only the final text. Frequent access to large language models (LLMs) during the attack process incurs significant costs and is often limited, as seen with the ChatGPT-4o model’s message limit of 50 per week~\cite{hayawi2024cross}. 
Given these constraints, it is essential to consider multi-task adversarial attacks in real-world scenarios, where the model feedback (i.e., the output text) reflects the black-box nature of the victim model. Furthermore, query limitations are a critical factor, as they optimize resource efficiency and minimize the risk of detection, making them a key consideration in practical attacks.
Therefore, this paper is guided by the following challenge:

\textit{How can attackers use \textbf{few-shot} queries to generate adversarial examples in a \textbf{black-box} multi-task learning model with \textbf{diverse task types}?}

In black-box and few-shot scenarios, a straightforward strategy is the transfer attack, where adversarial examples are generated in a substitute model without directly querying the victim model. However, this approach requires the availability of a well-trained substitute model. In the absence of such a model, attackers need to train a substitute model similar to the victim model. 
The transferability of adversarial examples depends on the similarity between these models, allowing adversarial examples crafted for the substitute to also affect the victim model.
Training a substitute model typically requires significant high-quality data, either from the victim model’s training set ($\mathbf{X}_{\text{train}}, \mathbf{Y}_{\text{train}}$) or by querying the victim model to generate predicting labels $\hat{\mathbf{Y}}_{\text{train}}$ for $\mathbf{X}_{\text{train}}$. However, acquiring such abundant data is often impractical, especially with query limitations in real-world settings.

Therefore, we propose the \textbf{deep-level attack hypothesis}:
adversarial examples generated from a substitute model trained with deep-level labels can transfer effectively across multiple downstream tasks. For instance, in a multi-task model for bird-vs-cat images, tasks like mammal classification, flying ability, and leg count can be tackled using deep-level labels like species classification, enabling adversarial examples to transfer across these tasks. Rather than training a substitute model to match the victim model, we focus on building a model with deep-level labels that can target all tasks with minimal data and few-shot queries, bypassing the need for large auxiliary datasets or extensive victim model queries.

Specifically, we propose \textbf{C}luster and \textbf{E}nsemble \textbf{M}ulti-task Text \textbf{A}dversarial Attack (CEMA), a framework that leverages a small set of easily accessible auxiliary data, such as victim texts. 
To obtain the deep-level labels, CEMA first vectorizes the input texts and their corresponding outputs into vector representations, and then concatenates the input and output vectors to form final representations. A binary clustering method is then applied to these representations to generate binary cluster labels, which serve as the deep-level labels. Subsequently, CEMA uses the auxiliary data-cluster label pairs to train a binary classification model as the substitute model, transforming the multi-task attack scenario into a single-task text classification attack. During the adversarial example generation phase, adversarial candidates are produced for each victim text. Simultaneously, CEMA randomly samples the auxiliary data-cluster label pairs and repeats the training process to generate multiple substitute models. The adversarial example that successfully attacks the majority of these substitute models is then selected as the final adversarial example.

To evaluate the performance of CEMA, we focus on the text classification and translation tasks within a multi-task learning framework.
Given the constraint of requiring only 100 queries, CEMA achieves a minimum improvement of 14.92\% in attack success rate (ASR) over the second-best method in the classification task. In the text translation task, CEMA improves the BLEU score by at least 0.05 compared to the second-best method.
With only 10 queries, CEMA still performs well, achieving an average ASR of 43.02\% and an average BLEU score of 0.22. On LLMs such as ChatGPT-4o~\cite{wang2025assessing} and Claude 3.5~\cite{bae2024enhancing}, CEMA achieves an average ASR of 36.15\% and an average BLEU score of 0.43. When the number of tasks is increased to six, CEMA also achieves great attack results. Meanwhile, CEMA can also attack the summarization, and text-to-image generation tasks.
The primary \textbf{contributions} are summarized as follows:

\begin{itemize}

    \item We define the deep-level attack hypothesis, demonstrating that adversarial examples generated from a substitute model trained with deep-level labels can effectively transfer across multiple downstream tasks.

\item We introduce CEMA, the first \textit{plug-and-play framework} that converts a multi-task attack into a classification attack. CEMA can craft adversarial examples with very limited queries and just final text outputs (Black-box attack). 

\item We empirically validate the effectiveness of CEMA across various victim models and datasets, consistently achieving state-of-the-art performance.
\end{itemize}

\section{Related Work}\label{related_work}
\subsection{Text Classification Adversarial Attack}
In  text adversarial research, the predominant methods revolve around scenarios with singular output results ~\cite{waghela2024modified,han2024bfs2adv,zhu2024limeattack,kang2024hybrid}. These studies focus on the techniques for morphing the original text into adversarial counterparts, including the manipulation of pivotal chars~\cite{ebrahimi2018Hotflip,gil2019white,ebrahimi2018adversarial,gao2018black,ren2019generating,jin2020Bert,li2019textbugger}, words~\cite{wang2022semattack,guo2021gradient,meng2020geometry,sato2018interpretable,cheng2019robust,lee2022query,li2020Bert,hu2024fasttextdodger,liu2024hqa,liu2023sspattack,li2019textbugger} and sentence. 
These methods are segmented into three distinct categories based on the response from the victim model, encompassing white-box attacks, soft-label black-box attacks, and hard-label black-box attacks. In white-box attacks, adversaries gain full access to all relevant information about the victim model. 

Hotflip~\cite{ebrahimi2018Hotflip} sequentially replaces key words based on their importance scores, while the FD method~\cite{papernot2016crafting} generates adversarial examples using model gradient information. In soft-label black-box attacks, many methods perturb words based on output probabilities~\cite{lee2022query,maheshwary2021strong,wang2021closer,li2020Bert}. Bert-ATTACK~\cite{li2020Bert} performs word-level attacks using a fine-tuned BERT; SememePSO~\cite{zang2020word} improves the search space for generating adversarial samples; Bae~\cite{garg2020Bae} replaces words via BERT; and DeepWordBug (DWB)~\cite{gao2018black} targets words prioritized by output probabilities.
Hard-label attacks reflect more realistic conditions. HLGA~\cite{maheshwary2021generating} uses stochastic initial words and a genetic algorithm to craft adversarial examples. HQA-attack~\cite{liu2024hqa} restores original words as much as possible to reduce disruption, then uses synonyms for the remaining changed words to enhance the attack.

\subsection{Neural Machine Translation Adversarial Attack}
Neural Machine Translation (NMT) models, which automatically convert input sentences into translated output, have achieved remarkable results by employing deep neural networks like Transformers ~\cite{bahdanau2014neural,vaswani2017attention}. These models are now extensively used across various applications due to their high performance. However, erroneous outputs generated by NMT models can lead to significant risks, particularly in security-sensitive contexts. 

Recent studies explore adversarial attacks on NMT models to address robustness concerns. Character-level NMT models are especially vulnerable to character-level manipulations like typos or word insertion/removal in black-box settings~\cite{belinkov2017synthetic,ebrahimi2018adversarial}. However, such manipulations are easily detectable by humans or review systems. Therefore, most attacks on NLP and NMT systems focus on word replacement.
Seq2sick~\cite{cheng2020seq2sick} uses a projected gradient method with group lasso and gradient regularization to perform non-overlapping and targeted keyword attacks. Similarly, Transfool~\cite{sadrizadeh2023transfool} applies gradient projection with a new optimization formulation and linguistic constraints to create fluent, semantically-preserving attacks. Morphin~\cite{tan2020s} perturbs inflections in clean examples to generate plausible and semantically similar adversaries. The kNN attack~\cite{michel2019evaluation} replaces words with their neighbors in the embedding space in a white-box, untargeted setting. RG~\cite{zou2019reinforced} adopts reinforcement learning to generate adversarial examples that preserve semantics and produce more reasonable tokens.

\subsection{Mutil-task Adversarial Attack}
A Multi-task Adversarial Attack is an adversarial machine learning strategy designed to generate examples that deceive multiple models or systems simultaneously~\cite{guo2020multi, Ghamizi_Cordy_Papadakis_Traon_2022}, rather than just one. As far as we know, there is currently no related work on multi-task adversarial attacks in the field of text.
In other fields, MTA~\cite{guo2020multi} is designed to generate adversarial perturbations for all three pre-trained classifiers simultaneously by leveraging shared knowledge among tasks. There is an attack method~\cite{sobh2021adversarial} that targets visual perception in autonomous driving, which is applied in a wide variety of multi-task
visual perception deep networks in distance estimation, semantic segmentation, motion detection, and object detection. MTADV~\cite{wang2024multi} is a multitask adversarial attack against facial authentication, which is effective against various facial data sets.

\subsection{Transfer Attack}
\textbf{Transfer attacks} leverage adversarial examples to target different models without requiring direct access, posing a significant security threat in black-box scenarios~\cite{papernot2017practical,dong2018boosting}.
Then, in the absence of a substitute model, several studies demonstrate that auxiliary data can also facilitate successful attacks through training a substitute model and leveraging transfer attacks~\cite{li2020practical,sun2022towards}. Additionally, more effective loss functions have been proposed to train substitute models~\cite{wang2021feature,li2020towards,naseer2019cross,richards2021adversarial,huan2020data}, as well as techniques to refine substitute models~\cite{xiaosen2023rethinking,yuan2021meta}.

\begin{figure*}[t]
  \centering \includegraphics[width=0.9\textwidth]{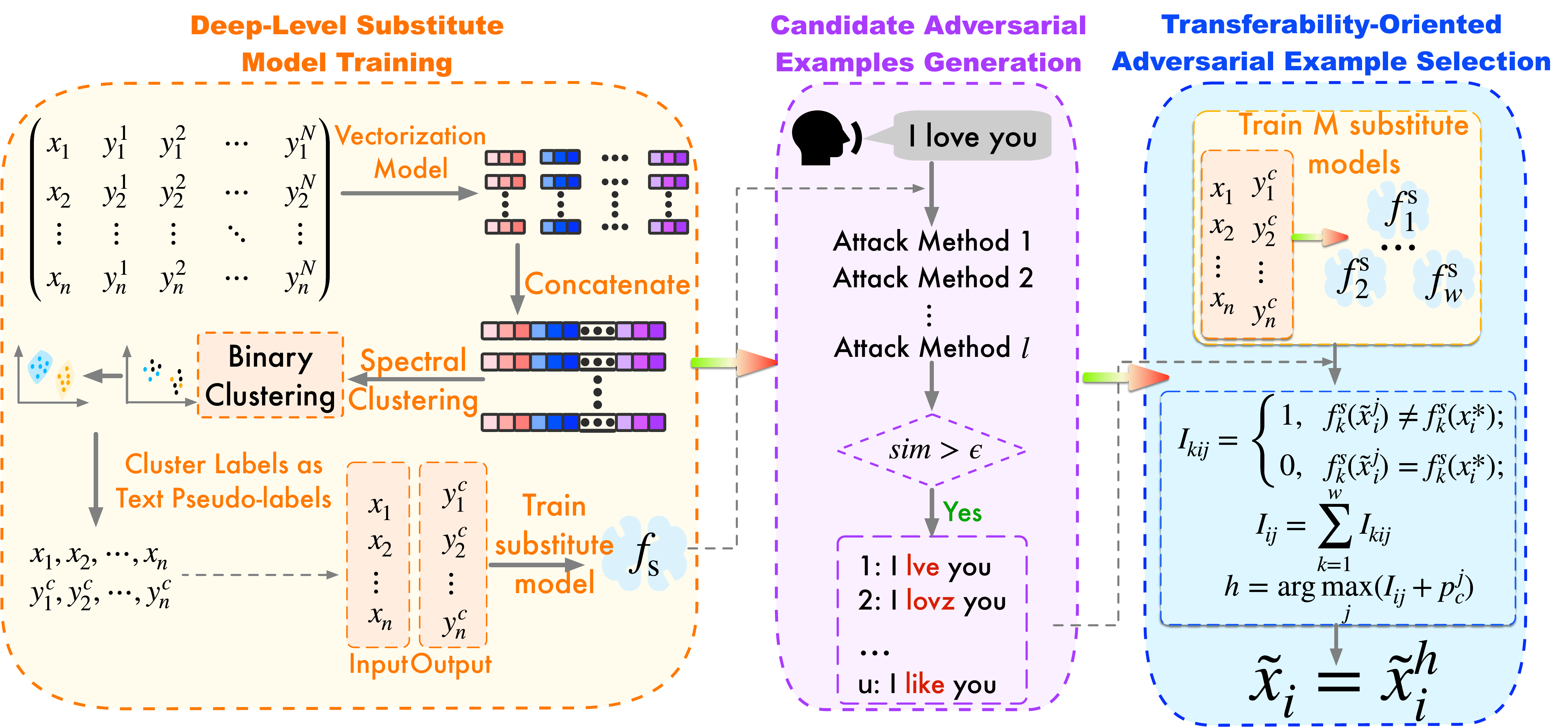} 
  \caption{\textbf{The  Overview  of CEMA.}
  \ding{182} During deep-level substitute model training, CEMA vectorizes the auxiliary data and labels using pre-trained models, concatenates the vectors, and applies binary clustering to generate deep-level labels, which are used to train the substitute model \( f_{\text{s}} \).
  \ding{183} For candidate adversarial example generation, \( l \) text classification attack methods are used to create \( l \) examples. Examples exceeding a similarity threshold \( \epsilon \) are selected as candidates, resulting in \( u \) candidates, where \( u < l \).
  \ding{184} CEMA trains \( w \)  substitute models using 80\% of the data pairs from the auxiliary text-cluster labels. The adversarial example attacking the most substitute models is chosen as the final adversarial example \( \tilde{x}_i \).
}\label{framework}
\vspace{-10pt}
\end{figure*}

\section{Preliminary}

\textbf{Transferability and multi-task learning.}
Transferability refers to the ability of adversarial examples, crafted for one model, to successfully attack other models~\cite{zhang2020two,papernot2017practical,mahmood2021beware,zhou2020dast}. 
Multi-task learning trains multiple related tasks together, enabling knowledge sharing and improving generalization~\cite{cai2025multimodal,pu2025integrating}.

\noindent\textbf{Auxiliary data.}
 Auxiliary data refers to additional information used to assist in generating adversarial examples. This data, which is not part of the original input, helps attackers better understand the target model's behavior or generate more effective adversarial inputs. For instance,  auxiliary data might include outputs from the victim model~\cite{chen2024reversible, papernot2017practical}.

\section{Method}

\subsection{Observations and Motivations}\label{o-a-m}

We consider a multi-task scenario with bird and cat images. Task 1 identifies if the image represents a mammal, Task 2 determines if the subject can fly, and Task 3 assesses the number of legs. Adversarial attacks on this multi-task model under black-box conditions with limited queries are challenging. However, by training a substitute model for bird-cat classification and perturbing it to misclassify a cat as a non-cat, we can modify the outputs of the three downstream tasks. Birds and cats share intrinsic species characteristics that are more fundamental and ``deep-level'' than the tasks of identifying mammalian status, flight capability, or legs' number. Based on this observation, we propose the \textbf{deep-level attack hypothesis}: 

\begin{assumption}[\textbf{Deep-level Attack Hypothesis}]
Adversarial examples $\tilde{x}$ of the victim text $x^*$ crafted using a substitute model $f_{\text{s}}$ trained with deep-level labels can effectively attack multiple downstream tasks in a multi-task model \( f_{\text{v}} \). Formally,
\begin{equation}
 f_{\text{s}}(x^*) \neq f_{\text{s}}(\tilde{x}),
\quad \text{and} \quad f_{\text{v}}(x^*) \neq f_{\text{v}}(\tilde{x}), 
\end{equation}
\end{assumption}

Based on the observations and hypothesis, we propose the \textbf{C}luster and \textbf{E}nsemble \textbf{M}ulti-task Text \textbf{A}dversarial Attack (CEMA) method, an overview of which is provided in Figure \ref{framework}. CEMA uses a deep-level substitute model to convert multi-task attack scenarios—characterized by varying numbers and types of tasks—into a single-task text classification attack scenario. Furthermore, CEMA employs multiple text classification attack methods to generate candidate adversarial examples and introduces a selection mechanism to identify the final adversarial example from the pool of candidates.
\subsection{Deep-level Substitute Model Training}
For auxiliary texts in auxiliary data, we query the victim model to obtain their corresponding outputs. 
These texts, along with their task-specific outputs, are then vectorized using a method such as a pre-trained model. The resulting vector representations are concatenated to form a unified representation. 
A binary clustering algorithm is applied to these vectorized representations, and cluster labels are assigned to each, which are treated as \textbf{deep-level labels} for both the text and its outputs.
Next, we train a substitute model, denoted as \( f_{\text{s}} \), using the auxiliary texts and their corresponding cluster labels. This substitute model \(f_{\text{s}} \) enables the transformation of the black-box multi-task attack scenario into a white-box text classification attack scenario. Moreover, \( f_{\text{s}} \) is treated as the substitute model trained with deep-level labels (cluster labels). 

\subsubsection{Representation Learning}
As discussed in Section \ref{o-a-m},  the deep-level label is derived from both the image and the outputs of three tasks: identifying whether the image represents a mammal, whether the subject can fly, and the number of legs. Thus, the deep-level label should simultaneously consider the impact of input data and the output data of the multi-task model. To achieve this, we vectorize the auxiliary texts and their outputs using a pre-trained model and concatenate them into a unified representation.

For each auxiliary text \( x_i \), we query the victim model \( f_{\text{v}} \) to obtain the corresponding outputs \( \{ y_i^1, y_i^2, \dots, y_i^N \} \), where \( N \) denotes the number of tasks. The pre-trained model \( f_e \) then vectorizes both \( x_i \) and its outputs, yielding vectors \( \{\boldsymbol{E}_{x_i}, \boldsymbol{E}_{y_i^1}, \dots, \boldsymbol{E}_{y_i^N}\} \). These vectors are concatenated to form the final vector \( \boldsymbol{E}_i \), which represents \( x_i \) and its corresponding outputs. Formally:
\begin{equation}
\begin{aligned}
\{ y_i^1, &y_i^2, \dots, y_i^N \}=f_{\text{v}}(x_i),\\
\boldsymbol{E}_{x_i} &= f_e(x_i), \quad
\boldsymbol{E}_{y_i^J} = f_e(y_i^J), \\
\boldsymbol{E}_{i} &= \text{Concat}(\boldsymbol{E}_{x_i},\boldsymbol{E}_{y_i^1},\dots,\boldsymbol{E}_{y_i^N}),
\end{aligned}
\end{equation}
where the $Concat$ indicates the concatenation of the vectors $\{\boldsymbol{E}_{x_i},\boldsymbol{E}_{y_i^1},\dots,\boldsymbol{E}_{y_i^N}\}$.

\subsubsection{Cluster Number and Substitute Model}
We set the number of clusters to 2, which can be interpreted as class \( C \) and its complement \( \overline{C} \). This choice ensures that the two clusters capture the most fundamental deep-level label, as the labels in three- or four-cluster configurations would ultimately merge into two. Experimental results in Section \ref{Ablation-Study} show that CEMA achieves optimal attack performance with two clusters. Attackers then employ the binary cluster method to obtain the cluster label \( y_i^{c} \) for auxiliary text \( x_i \). Finally, the binary substitute model \( f_{\text{s}} \) is trained using pairs of auxiliary text and cluster labels.  Formally:
\begin{equation}
\begin{aligned}
& \mathbf{D} = \{(x_1, y_1^{c}), (x_2, y_2^{c}), \dots, (x_n, y_n^{c})\}, \\
& \mathbf{\theta^*} = \arg\min_{\mathbf{\theta}} \frac{1}{|\mathbf{D}|} \sum_{(x_i, y_i^c) \in \mathbf{D}} (y_i^c - f_{\text{s}}(\mathbf{\theta}; x_i))^2,\\
\end{aligned}
\end{equation}
where \( y_i^c \) denotes the cluster label of \( E_i \), and \( \mathbf{\theta^*} \) represents the model parameters of the substitute model \( f_{\text{s}} \).

\subsection{Candidate Adversarial Examples Generation}\label{Can_Sample}
Once the substitute model \( f_{\text{s}} \) is trained, adversarial examples can be generated from it using any standard text classification method to attack the victim model \( f_{\text{v}} \) through transferability.
The perturbed text \( \tilde{x}_i \) of the victim texts $x_i^*$ is generated by:
\begin{equation}\label{Adv-attack}
\begin{split}
 & \tilde{x}_i  = \arg\max_{x_i} L(f_{\text{s}}(\tilde{x}_i),  f_{\text{s}}(x_i^*)), 
\quad \\   
\text{and} \quad f_{\text{v}}&(x_i^*)  \neq f_{\text{v}}(\tilde{x}_i), \quad \text{s.t.} \quad sim(\tilde{x}_i, x_i^*)  \geq \epsilon.
\end{split}
\end{equation}
where \( L(f(\tilde{x}_i), f(x_i)) \) is the loss function, \( f_{\text{s}}(\tilde{x}_i) \) represents the predictions of the substitute model for \( \tilde{x}_i \), and \( f_{\text{s}}(x_i^*) \) represents the predictions for \( x_i^* \). The term \( sim \) denotes the cosine similarity, which is calculated as $\frac{f_{\text{e}}(x_i^*) \cdot f_{\text{e}}(\tilde{x}_i)}{\|f_{\text{e}}(x_i^*)\| \cdot \|f_{\text{e}}(\tilde{x}_i)\|}$.
Finally, \( \epsilon \) is the similarity threshold. \textit{The condition \( sim(\tilde{x}_i, x_i^*) \geq \epsilon \) ensures that the adversarial example $\tilde{x}_i$ maintains high similarity with $x_i^*$ and retains its true label.}

Instead of relying on one attack method, we use multiple attack methods to generate diverse examples and establish criteria to select the most effective one. 
This strategy is motivated by two main reasons:  
(1) More adversarial examples increase the probability of at least one successfully attacking the victim model.  
(2) Diverse methods enhance the probability of surpassing the similarity threshold $\epsilon$ for multiple examples.
We apply $l$ attack methods to generate $l$ examples. The examples whose similarity exceeds the similarity threshold $\epsilon$ are identified as candidate adversarial examples ($\{ \tilde{x}^1_i, \tilde{x}^2_i, \dots, \tilde{x}^u_i \}$) for the victim text $x_i^*$, with $u \leq l$.
Denote the $p^j_i$ as the probability that $\tilde{x}^j_i$
can successfully attack the victim model. \( A^u \) denotes the event that at least one of the \( u \) attack methods attack the victim model, with the probability termed as $P(A^u)$. Denote $p_{ij}^{\text{sim}}$ as the probability that the similarity of $\tilde{x}_i^j$ and $x_i^*$ exceed the similarity threshold $\epsilon$. \( B^u \) denotes the event that at least one of the \( u \) adversarial examples exceed the similarity threshold $\epsilon$, with the probability termed as $P(B^u)$.
Reason (1) and (2) are mathematically expressed in Theorem \ref{max_prob2}.

\begin{theorem}\label{max_prob2}
$P(A^u)$ and $P(B^u)$ increase with $u$ increases . Formally:
\begin{equation}
\begin{array}{c}
P(A^u) \leq P(A^{u+1}), P(B^u) \leq P(B^{u+1})\quad \\ \text{for all} \quad u \geq 1,\\
\end{array}
\end{equation}
\end{theorem}

\textit{Proof.} Please see Section \ref{proof_max_prob2} in Appendix.

\subsection{Transferability-Oriented Adversarial Example Selection}\label{final_Sample}

After generating additional adversarial candidates in Section \ref{Can_Sample}, we select the one with the highest transferability. 
We achieve this by retraining multiple substitute models and evaluating the transferability of each candidate adversarial example based on the number of models it successfully attacks. The adversarial candidate that successfully attacks the greatest number of substitute models is selected as the final adversarial example. The detailed steps are as follows:
\ding{182} \textbf{Training  Substitute Models:}  
We randomly sample 80\% of the auxiliary text-cluster label pairs to form the training set for each  substitute model. This process is repeated \(w\) times, resulting in \(w\) substitute models \(\{f_1^{\text{s}}, f_2^{\text{s}}, \dots, f_w^{\text{s}}\}\).  \ding{183} \textbf{Calculating Transferability:}  
For each victim text \(x_i^*\) and its  adversarial candidates \( \{ \tilde{x}^1_i, \tilde{x}^2_i, \dots, \tilde{x}^u_i \} \). The transferability score for \(\tilde{x}^j_i\) is computed as follows:

\begin{equation}
I_{kij} = 
\begin{cases}
1, & \text{if } f_k^{\text{s}}(\tilde{x}_i^j) \neq f_k^{\text{s}}(x_i^*), \\
0, & \text{if } f_k^{\text{s}}(\tilde{x}_i^j) = f_k^{\text{s}}(x_i^*),
\end{cases}
\quad I_{ij} = \sum_{k=1}^w I_{kij}.
\end{equation}
where \( f_k^{\text{s}}\left(\tilde{x}^j_i\right) \) represents the output label of \( \tilde{x}^j_i \) produced by the substitute model \( f_k^{\text{s}} \), and \( f_k^{\text{s}}(x_i^*) \) is the output label of \( x_i^* \) from the $f_k^{\text{s}}$. If \( f_k^{\text{s}}\left(\tilde{x}^j_i\right) \neq f_k^{\text{s}}(x_i^*) \), then \( \tilde{x}^j_i \) successfully attacks \( f_k^{\text{s}} \). Thus, \( I_{i j} \) counts the number of  substitute models that \( \tilde{x}^j_i \) successfully attacks. The adversarial example attacking the most  substitute models is selected as the final adversarial example of the victim text $x_i^*$. If multiple candidate adversarial examples successfully attack the most  substitute models, the candidate with the largest change in output probability from the substitute model $f_{\text{s}}$  before and after the attack, is selected as the final adversarial example. Formally, this is expressed as:
\begin{equation}
\small
p_c^j=p_{\hat{y}}(x_i^{*}) - p_{\hat{y}}(\tilde{\mathbf{x}}_i^{j}),
h = \arg\max_j\ (I_{i j}+p_c^j),\tilde{x}_i=\tilde{x}^h_i,
\end{equation}
$\hat{y}$ denotes the predicted label of \( x_i^{*} \) under \( f_{\text{s}} \), with \( p_{\hat{y}}(x_i^{*}) \) and \( p_{\hat{y}}(\tilde{\mathbf{x}}_i^{j}) \) being the probabilities that \( x_i^{*} \) and \( \tilde{\mathbf{x}}_i^{j} \) are predicted as \( \hat{y} \), respectively. Since \( p_c^j \) represents probability variation, and the generated adversarial examples succeed in attacking \( f_{\text{s}} \), \( p_c^j \) lies in [0,1). When multiple adversarial examples attack the substitute model, \( h = \arg\max_j\ (I_{i j}+p_c^j) \) accounts for the probability variation in \( f_{\text{s}} \), with \( \tilde{x}_i \) being the final adversarial example.

\section{Experiment}
\subsection{Experiment Setup}
\textbf{Dataset:}
We use the \textbf{SST5} and \textbf{Emotion} datasets. Detailed statistics are in Appendix \ref{Details-of-Data}, Table \ref{stat}.

\noindent\textbf{Victim models and baselines in LLMs:} We conduct attacks on advanced LLMs, ChatGPT-4o~\cite{pang2024chatgpt} and Claude 3.5~\cite{bae2024enhancing}. We use prompt learning to build agents on LLMs as victim models. The prompt is: ``Translate the following text into French and Chinese, and predict the category label of the text, which can be one of the following: Sadness, Joy, Love, Anger, Fear, Surprise.''. 
As no existing research examines attacks on multi-task learning outputs of large models in a black-box setting, we adopt the approach of randomly removing the last character from 40\% of the words in a sentence as a baseline for comparison, which is recorded as Random-Del.

\begin{table*}[t]
\centering
\caption{The attack performance of LLMs.
}
\label{llm_attack}
\resizebox{0.8\textwidth}{!}{%
\begin{tabular}{@{}c|c|ccc|ccc|c@{}}
\toprule
\multirow{3}{*}{Data}    & Models     & \multicolumn{3}{c|}{ChatGPT-4o} & \multicolumn{3}{c|}{Claude 3.5} & \multirow{3}{*}{Total-Qry↓} \\ \cmidrule(lr){2-8}
                         & Tasks      & Classification  & En-Fr & En-Zh & Classification  & En-Fr & En-Zh &                          \\ \cmidrule(lr){2-8}
                         & Metrics    & ASR(\%)↑        & BLEU↓ & BLEU↓ & ASR(\%)↑        & BLEU↓ & BLEU↓ &                          \\ \midrule
\multirow{2}{*}{Emotion} & Random-Del & 14.20            & 0.47  & 0.64  & 17.65           & 0.44  & 0.52  & 0                        \\
                         & \cellcolor{lightpurple}\textbf{CEMA}       & \cellcolor{lightpurple}\textbf{32.05}           & \cellcolor{lightpurple}\textbf{0.39}  & \cellcolor{lightpurple}\textbf{0.33}  & \cellcolor{lightpurple}\textbf{36.80}            & \cellcolor{lightpurple}\textbf{0.38}  & \cellcolor{lightpurple}\textbf{0.35}  & \cellcolor{lightpurple}100                     \\ \midrule
\multirow{2}{*}{SST5}    & Random-Del & 15.33           & 0.64  & 0.66  & 16.27           & 0.62  & 0.54  & 0                        \\
                         & \cellcolor{lightpurple}\textbf{CEMA}       & \cellcolor{lightpurple}\textbf{38.63}           & \cellcolor{lightpurple}\textbf{0.32}  & \cellcolor{lightpurple}\textbf{0.27}  & \cellcolor{lightpurple}\textbf{37.12}           & \cellcolor{lightpurple}\textbf{0.29}  & \cellcolor{lightpurple}\textbf{0.25}  & \cellcolor{lightpurple}100                      \\ \bottomrule
\end{tabular}
}
\end{table*}

\noindent\textbf{Victim models and baselines in multi-model multi-task learning (M3TL):} 
M3TL trains multiple models to simultaneously solve various tasks~\cite{pineda2022multi,xiang2024multi,jin2021multi}. 
Inspired by M3TL, we combine several pre-trained models to simulate multi-task learning.
 For classification, we employ  dis-sst5, ro-sst5, dis-emotion, and ro-emotion models. For translation, we utilize the Opus-MT model for English-to-Chinese (En-Zh) task and the T5-small model for English-to-French (En-Fr) task. The model URLs are provided in Table \ref{url} in the Appendix.
To simulate a real-world attack scenario, we incorporate two commercial translation APIs: Baidu Translate (En-Fr)~\cite{he2015baidu} and Ali Translate (En-Zh)~\cite{wan2022alibaba}. Three multi-task victim models are created: \textbf{Victim Model A} (dis-sst5, dis-emo, opus-mt), \textbf{Victim Model B} (ro-sst5, ro-emo, t5-small), and \textbf{Victim Model C} (Baidu and Ali Translate).
Previous multi-task attack methods~\cite{liu2017adversarial} require extensive queries and focus on white-box models with similar tasks, making them ineffective against multi-task black-box models. In comparison, CEMA targets both classification and translation tasks with fewer queries. For single-task attacks, we compare CEMA with methods for text classification (BAE~\cite{garg2020Bae}, FD~\cite{papernot2016crafting}, Hotflip~\cite{ebrahimi2018Hotflip}, SememePSO~\cite{zang2020word}, TextBugger~\cite{ren2019generating}) and translation (Hot-trans~\cite{ebrahimi2018Hotflip}, kNN~\cite{michel2019evaluation}, Morphin~\cite{tan2020s}, RA~\cite{zou2019reinforced}, Seq2Sick~\cite{cheng2020seq2sick}, TransFool~\cite{sadrizadeh2023transfool}). All methods are limited to 55 queries per victim text to ensure a fair comparison. Details are provided in Tables \ref{baseline-Class} and \ref{baseline-Trans} in Appendix \ref{Details-of-Baseline}.

\noindent\textbf{Metrics, substitute model, clustering and vectorization methods:} 
We evaluate the following metrics: \ding{182} ASR: A higher ASR indicates better attack efficacy. \ding{183} Total Queries (Total-Qry): Fewer queries reflect a more efficient attack. \ding{184} BLEU: A lower BLEU score indicates a more effective attack result~\cite{lateef2024evaluation}. 
\ding{185} Similarity~(Sim.): A higher Sim. score indicates a more effective attack result.
Details of the substitute model are provided in Section \ref{substitute-model-architecture} of the Appendix. We train six substitute models on a server equipped with a 24GB NVIDIA 3090 GPU, with a training time of approximately 4 minutes per model due to limited auxiliary data. Each model has a size of 418 MB. Spectral clustering~\cite{ng2001spectral} is used for clustering, and mT5~\cite{xue2020mt5} is used for vectorization.

\noindent\textbf{Applied attack methods, auxiliary data, and hyperparameter:}
We apply a combination of attack methods, including Hotflip~\cite{ebrahimi2018Hotflip}, FD~\cite{papernot2016crafting}, and TextBugger~\cite{ren2019generating}. These methods assume that the attacker has access to the victim texts during the attack, indicating that the victim texts are readily available for modification. For auxiliary data, we select 100 unlabeled victim texts. The similarity threshold \( \epsilon \) is set to 0.8.

\subsection{Main Attack Results of LLMs and M3TL}
As shown in Table \ref{llm_attack},  CEMA achieves substantial attack effectiveness on LLMs, with a best ASR of 38.63\% and a best BLEU score of 0.25.
 Tables \ref{main-table} and \ref{VM-C} highlight that CEMA achieves state-of-the-art (SOTA) performance on the SST5 and Emotion datasets across victim models A, B, and C. In a black-box setting with 100 queries, CEMA attains an ASR exceeding 59\%, reaching up to 80.80\% in classification tasks. For translation tasks, CEMA achieves a BLEU score below 0.16, outperforming the second-best method. Against victim model C (Baidu and Ali Translate) using only 100 auxiliary texts, CEMA exceeds Morphin and TransFool, attaining BLEU scores under 0.35 with just 100 queries. 
\begin{table*}[t]
\centering
\caption{The attack performance of CEMA.
Text classification tasks use the ASR(\%)↑ metric, while text translation tasks use the BLEU↓ metric. Other adversarial attack methods can only be applied to their specific tasks, whereas CEMA simultaneously attacks all tasks.}
\label{main-table}
\resizebox{1\textwidth}{!}{%
\begin{tabular}{@{}c|cccccc|cccccc@{}}
\toprule
Dataset             & \multicolumn{6}{c|}{SST5}                                                                                            & \multicolumn{6}{c}{Emotion}                                                                                          \\ \midrule \midrule
Victim Model        & \multicolumn{3}{c|}{Victim Model A}                                 & \multicolumn{3}{c|}{Victim Model B}            & \multicolumn{3}{c|}{Victim Model A}                                 & \multicolumn{3}{c}{Victim Model B}             \\ \midrule
Text Classification & \multicolumn{3}{c|}{dis-sst5 (A)}                                   & \multicolumn{3}{c|}{ro-sst5 (B)}               & \multicolumn{3}{c|}{dis-emotion (A)}                                & \multicolumn{3}{c}{ro-emotion (B)}             \\ \midrule
Metric              & ASR(\%)↑       & Sim.↑          & \multicolumn{1}{c|}{Total-Qry↓}   & ASR(\%)↑       & Sim.↑          & Total-Qry↓   & ASR(\%)↑       & Sim.↑          & \multicolumn{1}{c|}{Total-Qry↓}   & ASR(\%)↑       & Sim.↑          & Total-Qry↓   \\ \midrule
Bae                 & 42.71          & 0.888          & \multicolumn{1}{c|}{47360}        & 39.14          & 0.887          & 47471        & 31.55          & 0.925          & \multicolumn{1}{c|}{59626}        & 28.50          & 0.924          & 55935        \\
FD                  & 25.20          & 0.939          & \multicolumn{1}{c|}{27758}        & 22.30          & \textbf{0.982} & 21459        & 47.10          & {\ul 0.948}    & \multicolumn{1}{c|}{66035}        & 20.75          & {\ul 0.979}    & 26719        \\
Hotflip             & 41.50          & 0.951          & \multicolumn{1}{c|}{25459}        & 29.03          & 0.951          & 25945        & 46.85          & 0.942          & \multicolumn{1}{c|}{21658}        & 41.65          & 0.952          & 22409        \\
PSO                 & 45.14          & {\ul 0.954}    & \multicolumn{1}{c|}{24398}        & 41.50          & 0.954          & 27360        & 46.05          & 0.945          & \multicolumn{1}{c|}{19713}        & 44.95          & 0.964          & 19757        \\
TextBugger          & 30.36          & \textbf{0.978} & \multicolumn{1}{c|}{69527}        & 20.85          & {\ul 0.978}    & 67007        & 35.10          & \textbf{0.981} & \multicolumn{1}{c|}{25216}        & 29.40          & \textbf{0.981} & 25128        \\
Leap                & 32.55          & 0.953          & \multicolumn{1}{c|}{21548}        & 30.07          & 0.944          & 21083        & 26.30          & 0.934          & \multicolumn{1}{c|}{15492}        & 15.50          & 0.939          & 15315        \\
CT-GAT              & 29.37          & 0.939          & \multicolumn{1}{c|}{46233}        & 24.80          & 0.926          & 82963        & 25.90          & 0.916          & \multicolumn{1}{c|}{47338}        & 26.75          & 0.927          & 47139        \\
HQA                 & 46.11          & 0.936          & \multicolumn{1}{c|}{64864}        & 39.64          & 0.929          & 64267        & 37.35          & 0.934          & \multicolumn{1}{c|}{65725}        & 35.85          & 0.925          & 47449        \\ \midrule
CEMA                & \textbf{73.57} & 0.934          & \multicolumn{1}{c|}{\textbf{100}} & \textbf{75.66} & 0.927          & \textbf{100} & \textbf{80.80} & 0.926          & \multicolumn{1}{c|}{\textbf{100}} & \textbf{60.40} & 0.931          & \textbf{100} \\ \midrule
Text Classification & \multicolumn{3}{c|}{dis-emotion (A)}                                & \multicolumn{3}{c|}{ro-emotion (B)}            & \multicolumn{3}{c|}{dis-emotion (A)}                                & \multicolumn{3}{c}{ro-emotion (B)}             \\ \midrule
Metric              & ASR(\%)↑       & Sim.↑          & \multicolumn{1}{c|}{Total-Qry↓}   & ASR(\%)↑       & Sim.↑          & Total-Qry↓   & ASR(\%)↑       & Sim.↑          & \multicolumn{1}{c|}{Total-Qry↓}   & ASR(\%)↑       & Sim.↑          & Total-Qry↓   \\ \midrule
Bae                 & 39.81          & 0.894          & \multicolumn{1}{c|}{60399}        & 14.65          & 0.896          & 62013        & 32.25          & 0.926          & \multicolumn{1}{c|}{48266}        & 32.95          & 0.923          & 48244        \\
FD                  & 35.43          & 0.921          & \multicolumn{1}{c|}{64576}        & 9.55           & 0.934          & 36553        & 22.30          & 0.932          & \multicolumn{1}{c|}{28310}        & 17.50          & \textbf{0.982} & 40730        \\
Hotflip             & 33.39          & 0.943          & \multicolumn{1}{c|}{24001}        & 22.80          & 0.946          & 27139        & 29.00          & 0.949          & \multicolumn{1}{c|}{31559}        & 28.05          & 0.949          & 31824        \\
PSO                 & 41.90          & 0.968          & \multicolumn{1}{c|}{19934}        & 35.25          & 0.940          & 20885        & 39.50          & {\ul 0.952}    & \multicolumn{1}{c|}{26144}        & 37.65          & 0.951          & 26741        \\
TextBugger          & 30.00          & \textbf{0.972} & \multicolumn{1}{c|}{25084}        & 40.95          & {\ul 0.986}    & 25084        & 20.85          & \textbf{0.978} & \multicolumn{1}{c|}{67007}        & 21.45          & {\ul 0.978}    & 67029        \\
Leap                & 21.00          & {\ul 0.968}    & \multicolumn{1}{c|}{15315}        & 26.00          & {\ul 0.947}    & 15492        & 40.55          & 0.926          & \multicolumn{1}{c|}{21503}        & 37.65          & 0.911          & 21614        \\
CT-GAT              & 39.32          & 0.927          & \multicolumn{1}{c|}{47206}        & 33.45          & 0.924          & 47493        & 28.10          & 0.904          & \multicolumn{1}{c|}{57593}        & 30.85          & 0.906          & 56001        \\
HQA                 & 37.76          & 0.945          & \multicolumn{1}{c|}{47382}        & 31.95          & 0.931          & 65062        & 37.40          & 0.912          & \multicolumn{1}{c|}{49592}        & 36.40          & 0.911          & 51184        \\ \midrule
CEMA                & \textbf{62.27} & 0.934          & \multicolumn{1}{c|}{\textbf{100}} & \textbf{64.01} & 0.927          & \textbf{100} & \textbf{65.40} & 0.926          & \multicolumn{1}{c|}{\textbf{100}} & \textbf{59.60} & 0.931          & \textbf{100} \\ \midrule
Text Translation    & \multicolumn{3}{c|}{opus-mt(En-Zh) (A)}                             & \multicolumn{3}{c|}{t5-small(En-Fr) (B)}       & \multicolumn{3}{c|}{opus-mt(En-Zh) (A)}                             & \multicolumn{3}{c}{t5-small(En-Fr) (B)}        \\ \midrule
Metric              & BLEU↓          & Sim.↑          & \multicolumn{1}{c|}{Total-Qry↓}   & BLEU↓          & Sim.↑          & Total-Qry↓   & BLEU↓          & Sim.↑          & \multicolumn{1}{c|}{Total-Qry↓}   & BLEU↓          & Sim.↑          & Total-Qry↓   \\ \midrule
Hot-trans           & 0.24           & 0.846          & \multicolumn{1}{c|}{21570}        & 0.24           & 0.842          & 20885        & 0.20           & 0.859          & \multicolumn{1}{c|}{20686}        & 0.19           & 0.854          & 21680        \\
KNN                 & 0.31           & 0.873          & \multicolumn{1}{c|}{13680}        & 0.31           & 0.883          & 13680        & 0.61           & 0.935          & \multicolumn{1}{c|}{29481}        & 0.28           & 0.906          & 13437        \\
Morphin             & 0.30           & 0.894          & \multicolumn{1}{c|}{15006}        & 0.37           & 0.907          & 24531        & 0.27           & 0.869          & \multicolumn{1}{c|}{11183}        & 0.22           & 0.887          & 8486         \\
RA                  & 0.25           & 0.872          & \multicolumn{1}{c|}{7028}         & 0.19           & 0.865          & 9415         & 0.23           & 0.852          & \multicolumn{1}{c|}{6166}         & 0.21           & 0.865          & 4663         \\
Seq2sick            & 0.38           & 0.881          & \multicolumn{1}{c|}{9835}         & 0.46           & {\ul 0.926}    & 13371        & 0.62           & {\ul 0.945}    & \multicolumn{1}{c|}{15669}        & 0.29           & 0.892          & 8951         \\
TransFool           & 0.77           & \textbf{0.949} & \multicolumn{1}{c|}{7337}         & 0.44           & 0.894          & 8641         & 0.81           & \textbf{0.962} & \multicolumn{1}{c|}{8597}         & 0.67           & {\ul 0.924}    & 7912         \\ \midrule
CEMA                & \textbf{0.14}  & {\ul 0.934}    & \multicolumn{1}{c|}{\textbf{100}} & \textbf{0.18}  & \textbf{0.927} & \textbf{100} & \textbf{0.15}  & 0.926          & \multicolumn{1}{c|}{\textbf{100}} & \textbf{0.23}  & \textbf{0.931} & \textbf{100} \\ \bottomrule
\end{tabular}
}
\end{table*}

\begin{table}[t!]
\centering
\caption{Attack performance of different methods on victim model C, which consists of two commercial closed-source translation APIs,  Alibaba  and Baidu Translate.}\label{VM-C}
\resizebox{0.49\textwidth}{!}{%
\begin{tabular}{@{}c|c|ccc|ccc@{}}
\toprule
\multirow{2}{*}{Dataset} & Victim Model C & \multicolumn{3}{c|}{Baidu Translate (en-fr)(C)} & \multicolumn{3}{c}{Ali Translate (en-zh)(C)}  \\ \cmidrule(l){2-8} 
                         & Method         & BLEU↓          & Sim.↑           & Total-Qry↓   & BLEU↓         & Sim.↑          & Total-Qry↓   \\ \midrule \midrule
\multirow{3}{*}{SST5}    & Morphin        & 0.54           & 0.904           & 89461        & 0.60          & 0.931          & 107075       \\
                         & TransFool      & 0.51           & 0.921           & 52001        & 0.59          & 0.928          & 68952        \\ \cmidrule(l){2-8} 
                         & CEMA           & \textbf{0.29}  & \textbf{0.934}  & \textbf{100} & \textbf{0.15} & \textbf{0.934} & \textbf{100} \\ \midrule
\multirow{3}{*}{Emotion} & Morphin        & 0.40           & 0.897           & 55581        & 0.55          & 0.915          & 25473        \\
                         & TransFool      & 0.36           & 0.903           & 25416        & 0.49          & 0.923          & 61820        \\ \cmidrule(l){2-8} 
                         & CEMA           & \textbf{0.35}  & \textbf{0.931}  & \textbf{100} & \textbf{0.29} & \textbf{0.931} & \textbf{100} \\ \bottomrule
\end{tabular}
}
\end{table}

\subsection{Ablation Study}\label{Ablation-Study}
To investigate factors influencing CEMA attack performance, we conduct an ablation study on the number of clusters, attack methods, clustering, and vectorization. We experiment with two to four clusters and increase attack methods from one to three, using TextBugger alone and a combination of Hotflip, FD, and TextBugger. For clustering, we use spectral clustering as the main method, along with K-means~\cite{krishna1999genetic} and BIRCH~\cite{zhang1996birch}. For vectorization, we primarily use mT5~\cite{xue2020mt5}, with XLM-R~\cite{conneau2019unsupervised} and one-hot encoding~\cite{rodriguez2018beyond} as alternatives.
One-hot encoding transforms categorical data into binary vectors without requiring access to any external data,  preventing data leakage in pre-trained process. Given the high computational cost of utilizing large models, we primarily conduct ablation experiments on victim models A, B, and C.

\begin{figure}[t]
\centering
\includegraphics[width=0.45\textwidth]{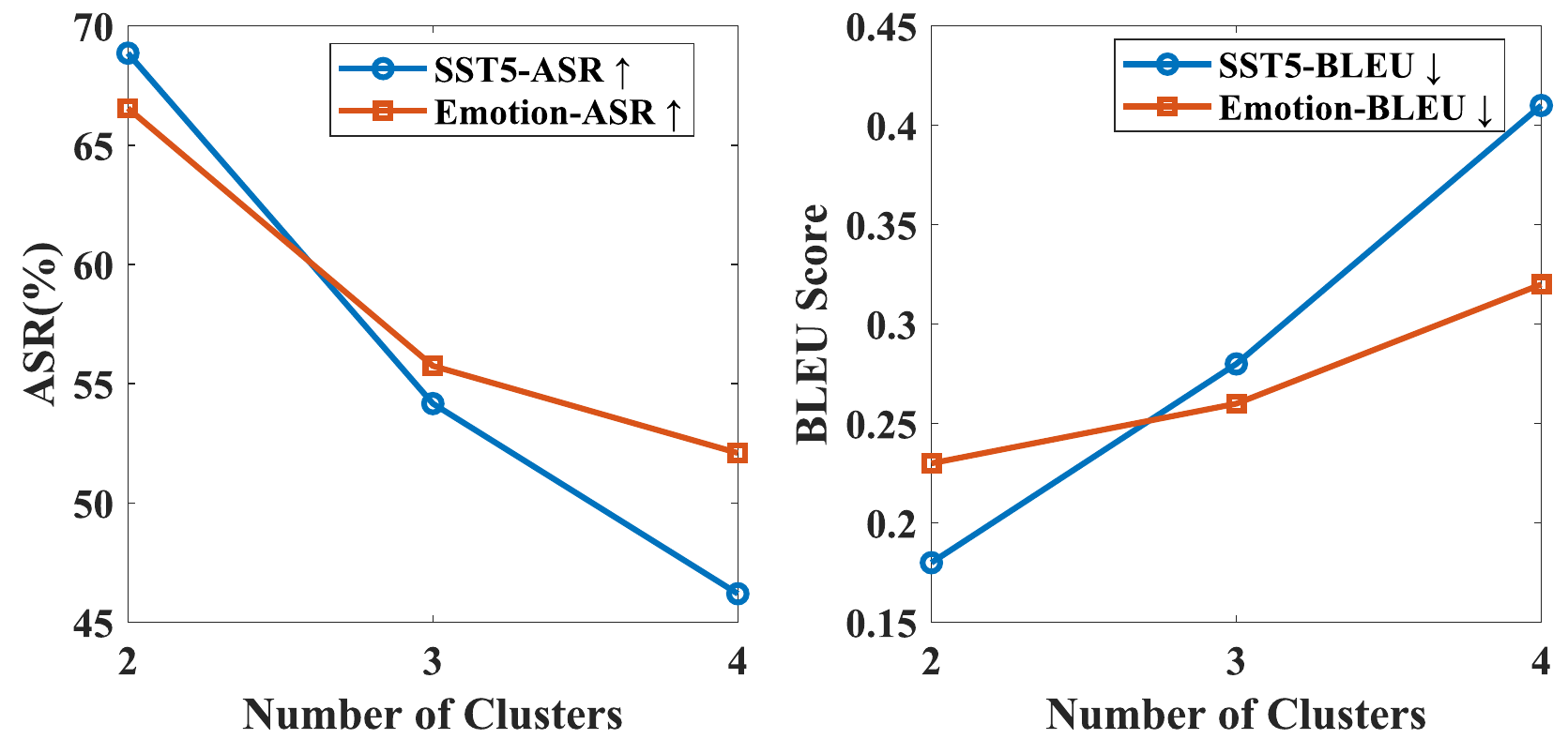}
\caption{The average ASR and BLUE of different clusters' number. Fewer clusters result in the better attack results.}
\label{c-num}
\end{figure}

\noindent\textbf{Cluster number and attack method number.}
As shown in Figure \ref{c-num}, increasing the number of clusters from 2 to 4 reduces attack performance. The average ASR drops from 58.83\% and 64.55\% to 46.20\% and 52.10\%, while BLEU scores increase from 0.16 and 0.18 to 0.41 and 0.32. \textit{The best attack performance is achieved with two clusters}. Additionally, as shown in Table \ref{Example-Number} in the Appendix, reducing the number of attack methods lowers ASR and raises BLEU scores, as smaller adversarial spaces reduce attack effectiveness. Increasing the number of attack algorithms from one to three improves average ASR by 30.3\% while decreasing BLEU by 0.16, indicating that more attack methods enhance performance.

\begin{figure}[t]
    \centering
    \includegraphics[width=0.50\textwidth]{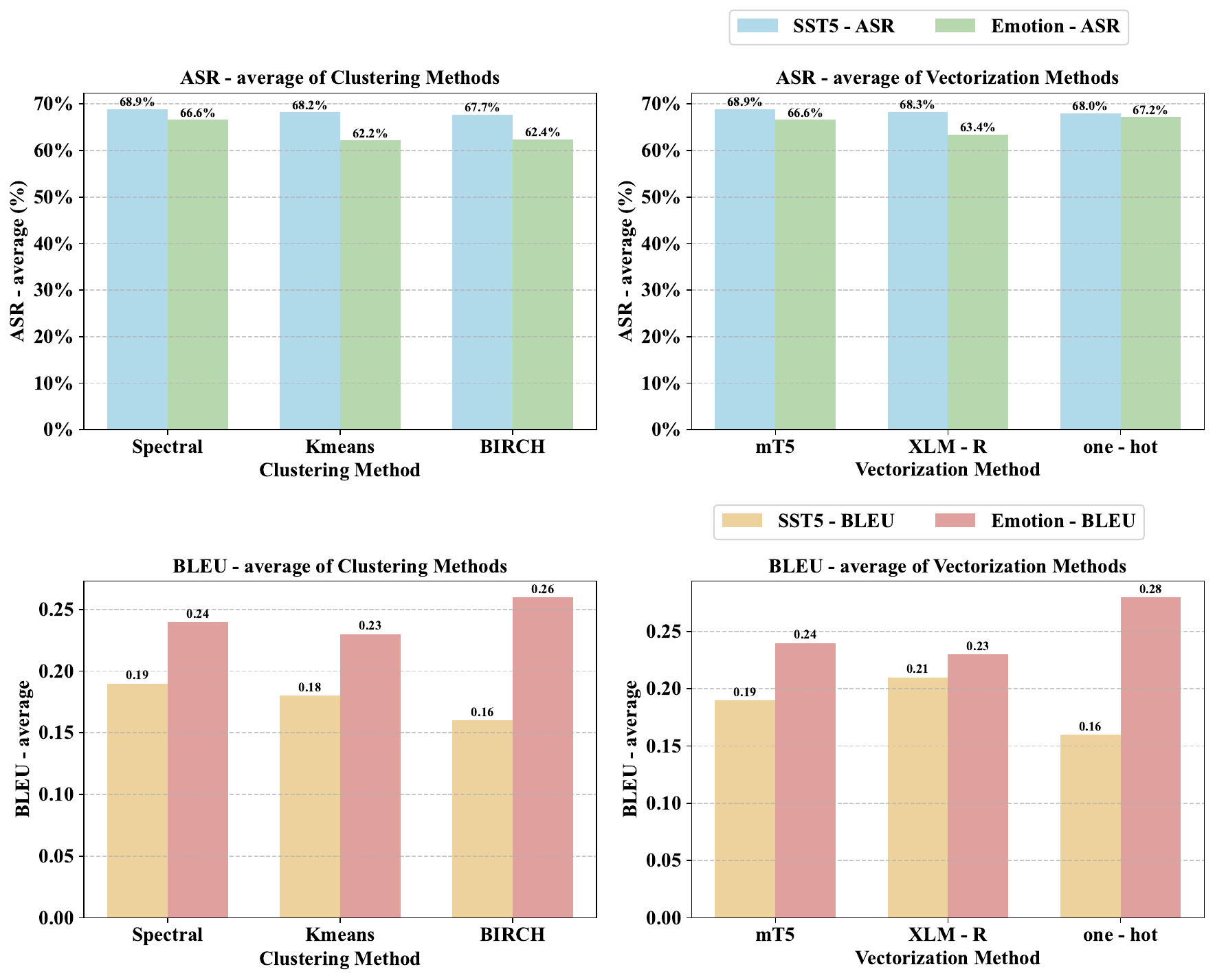} 
    \caption{The average ASR and BLUE of CEMA under various clustering and vectorization methods.} 
    \label{c-v-method} 
    \vspace{-15pt}
\end{figure}

\noindent\textbf{Clustering and vectorization methods.}
As shown in Figure \ref{c-v-method} 
, both clustering and vectorization methods have random effects on attack performance. The average ASR for Spectral, KMeans, and BIRCH are 67.7\%, 65.2\%, and 65.1\%, respectively, while BLEU scores are approximately 0.21, with no method consistently outperforming the others. This indicates that the impact of clustering methods on attack results is random. Similarly, the choice of vectorization methods leads to slight fluctuations in both classification and translation, but none achieves SOTA performance. The average ASR for mT5, XLM-R, and one-hot encoding is around 67.7\%, with BLEU scores ranging from 0.21 to 0.22, indicating the random impact of vectorization methods.

In summary, clustering and vectorization methods introduce randomness, and while increasing the number of attack methods improves effectiveness, it also prolongs attack time.

\subsection{Scalability of CEMA for More Numbers and Types of Tasks}
CEMA is extended to six downstream tasks, comprising four classification tasks and two translation tasks, as detailed in Table \ref{6-tasks} in the Appendix. It achieves an ASR of over 60\%, with BLEU scores below 0.3, demonstrating its effectiveness in six downstream tasks. 
To explore CEMA's scalability further, we apply it to a broader set of tasks using the Pokemon dataset~\cite{2022pokemon}
as victim data.
The victim models used include Google Translate~\cite{radford2021learning}, distilbart-cnn~\cite{sakhapara2022summarization}, and Stable Diffusion V2, with tasks such as translation~\cite{kol2018google}, summarization, and text-to-image generation.  The evaluation metrics include BLEU, ROUGE Drop Percentage (RDP)~\cite{lin2004rouge}, and CLIP Drop Percentage (CDP)~\cite{radford2021learning}. As shown in Table \ref{t2i} in the Appendix, CEMA exhibits strong performance across all tasks, suggesting its potential for extension. Specifically, the BLEU score is 0.29, with RDP and CDP scores of 47\% and 56\%, respectively. These results indicate that CEMA is well-suited for multi-task models involving a broader range of tasks and task types.

\subsection{Few-shot and Many-shot Queries}\label{few-and-add}
We evaluate CEMA’s attack performance under few-shot and many-shot queries, with query numbers set to 10, 50, 100, 1000, and 2000. The results, shown in Table \ref{few-show-results} in the Appendix, reveal that attack effectiveness improves with more queries. Notably, CEMA achieves an ASR of over 30\% and a BLEU score of 0.19 with only 10 queries.

\subsection{
Other Results}

\ding{182} we collect relevant internet data as auxiliary information for the victim texts, as described in Table \ref{data-data} (Appendix) and Section \ref{Low-quality}. CEMA achieves strong results, showing that attackers only need the victim texts' related information to successfully attack the multi-task system.
\ding{183} \textbf{Attack results under defense methods}
We explore defensive strategies against CEMA, including language modifiers and adversarial training. Detailed results are presented in Section \ref{def}. While these defenses reduce CEMA's effectiveness, it still maintains a significant level of attack efficacy.
\section{Limitation}
CEMA requires training multiple models and generating several candidate adversarial examples, which demands additional time and storage space.

\section{Conclusion}
We present a multi-task learning attack scenario where attackers have limited access to a model, only being able to query black-box outputs. Therefore, we propose the CEMA method, which achieves SOTA performance with just 100 queries and black-box outputs. Additionally, CEMA can integrate any text classification attack algorithm, and its performance improves as more attack methods are included. We plan to extend CEMA to multi-task models in other domains in the future.

\section{Acknowledgments}
This work was supported by the Fundamental Research Funds for the Central Universities, Sun Yat-sen University under Grants No. 23xkjc010, Shenzhen Science and Technology Program (No.KQTD20221101093559018), and National Natural Science Foundation of China (No. 62441619).

\bibliography{custom}

\appendix

\newpage
\section*{Overview of the Appendix}
This appendix includes our supplementary materials as follows:

- More details of the substitute model in Section \ref{substitute-model-architecture}

- More details of data in Section \ref{Details-of-Data}

- Url of victim model used in Section \ref{url of vm}

- Details of baselines in Section \ref{Details-of-Baseline}

- More details of defense methods in Section \ref{def}

- Definition of text classification adversarial examples and NMT adversarial examples in Section \ref{def of text clas}

- Experiment result for verifying independence in section \ref{verify}

- The proof of Theorem \ref{max_prob2} in Section \ref{proof_max_prob2}

- More Explanation for the Non-Independent Case in Section \ref{Non-Independent}

- More Details of M3TL in Section \ref{M3TL}

- Experiment Result under Low-quality Auxiliary Data Setting in Section \ref{Low-quality}
\newpage

\section{Substitute Model}\label{substitute-model-architecture}
\subsection{Substitute Model Architecture}
Our substitute model comprises 12 transformer blocks, each with 768 hidden units and 12 self-attention heads. Each transformer block consists of the following substructures:

\begin{itemize} 
\item \textbf{Self-Attention Layer:} The hidden size of the self-attention layer is 768.
\item \textbf{Position-wise Feed-Forward Network:} The network first projects the output of the attention layer to a 3072-dimensional space using a fully connected layer, followed by a ReLU activation for non-linearity, and finally projects the 3072-dimensional space back to a 768-dimensional space via another fully connected layer.

\item \textbf{Layer Normalization and Residual Connection:}
\begin{itemize}
    \item \textbf{Layer Normalization:} Applied to the output of each sub-layer to stabilize training.
    \item \textbf{Residual Connection:} Adds the normalized output to the input of the sub-layer.
\end{itemize}
\end{itemize}

\subsection{Substitute Model Training}
We provide a detailed description of the training of the substitute model with the transformer-based architecture. This substitute model consists of $12$ hidden layers with a dimensionality of $768$. The activation function ``GELU'' is used, The dropout rate is $0.4$.
The training process is optimized with the AdamW optimizer~\cite{yao2021adahessian}, with batch size set to $64$ and learning rate set to $6e-3$, over $5$ epochs.

\subsection{Computation Overhead of the Substitute Model Training}\label{train}

We train six substitute models on a server equipped with a 24 GB NVIDIA 3090 GPU. 
The training time for a single model is approximately 4 minutes, and the size of each trained model is 418 MB.

\section{Details of Datasets}\label{Details-of-Data}
\begin{table*}[t]
\centering
\caption{The statistics of datasets.}
\label{stat}
\resizebox{0.8\textwidth}{!}{
\begin{tabular}{@{}ccc@{}}
\toprule
Dataset & Classes & Labels' name                                              \\ \midrule \midrule
SST5    & 5       & Very positive, Positive, Neutral, Negative, Very negative \\
Emotion & 6       & Sadness, Joy, Love, Anger, Fear, Surprise                 \\ \bottomrule
\end{tabular}
}
\end{table*}
The details of datasets are presented in Table \ref{stat}
\section{The URLs of the Victim Models}
\label{url of vm}
The URLs of the victim models are presented in Table \ref{url}
\begin{table*}[t]
\centering
\caption{The URL of the Victim Models}\label{url}
\resizebox{0.85\textwidth}{!}{
\begin{tabular}{@{}c|c@{}}
\toprule
Model & Url \\ \midrule
dis-sst5(A) & \url{https://huggingface.co/SetFit/distilbert-base-uncased__sst5__all-train} \\
dis-emotion(A) & \url{https://huggingface.co/bhadresh-savani/distilbert-base-uncased-emotion} \\
opus-mt(En-Zh)(A) & \url{https://huggingface.co/Helsinki-NLP/opus-mt-En-Zh} \\
ro-sst5(B) & \url{https://huggingface.co/Unso/roberta-large-finetuned-sst5} \\
ro-emotion(B) & \url{https://huggingface.co/SamLowe/roberta-base-go_emotions} \\
t5-small(En-Fr)(B) & \url{https://huggingface.co/Alexle/T5-small-En-Fr} \\
Baidu Translate (En-Fr) (C)  & \url{https://api.fanyi.baidu.com/} \\
Ali Translate (En-Zh) (C)  & \url{https://translate.alibaba.com/} \\  \bottomrule
\end{tabular}
}
\end{table*}

\section{Details of Baselines}\label{Details-of-Baseline}
The details of baselines are presented in Table \ref{baseline-Class} and Table \ref{baseline-Trans}

\begin{table*}[t]
   \centering
   \caption{The details of the methods employed in the baseline comparisons. The Perturbed Level indicates the target of the attack methods, where ``word'' denotes the specific words targeted for perturbation, and ``char'' refers to the characters within a word that are altered by the attack method.}
   \label{baseline-combined}
   
   \begin{minipage}{\textwidth}
       \centering
       \subcaption{Information on the classification attack method used as the baseline.}
       \label{baseline-Class}
       \resizebox{0.9\textwidth}{!}{%
       \begin{tabular}{@{}cccccc@{}}
       \toprule
       Methods      & Perturbed Level & Gradient       & Soft-labels    & Hard-labels    & Knowledge \\ \midrule
       Bae          & Word            & \ding{55}   & \ding{51} & \ding{51} & black-box \\
       FD           & Char            & \ding{51} & \ding{51} & \ding{51} & white-box \\
       Hotflip      & Char            & \ding{51} & \ding{51} & \ding{51} & white-box \\
       PSO          & Word            & \ding{55}   & \ding{51} & \ding{51} & black-box \\
       TextBugger   & Char+Word       & \ding{51} & \ding{51} & \ding{51} & white-box \\
       Leap         & Word            & \ding{55}   & \ding{51} & \ding{51} & black-box \\
       CT-GAT       & Word            & \ding{55}   & \ding{51} & \ding{51} & black-box \\
       HQA          & Word            & \ding{55}   & \ding{51} & \ding{51} & black-box \\ \midrule
       CEMA         & Char+Word       & \ding{55}   & \ding{55}   & \ding{51} & black-box \\ \bottomrule
       \end{tabular}%
       }
   \end{minipage}
   
   \vspace{1em}
   
   \begin{minipage}{\textwidth}
       \centering
       \subcaption{Information on the translation attack method used as the baseline.}
       \label{baseline-Trans}
       \resizebox{0.9\textwidth}{!}{%
       \begin{tabular}{@{}cccccc@{}}
       \toprule
       Methods      & Perturbed Level & Gradient       & Soft-labels    & Hard-labels    & Knowledge \\ \midrule
       Hot-trans    & Char            & \ding{51} & \ding{55} & \ding{55} & white-box \\
       kNN          & Word            & \ding{51} & \ding{55} & \ding{55} & white-box \\
       Morphin      & Word            & \ding{55}   & \ding{55} & \ding{51} & black-box \\
       RA           & Word            & \ding{51} & \ding{55} & \ding{55} & white-box \\
       Seq2Sick     & Word            & \ding{51} & \ding{55} & \ding{51} & white-box \\
       TransFool    & Word            & \ding{55}   & \ding{55} & \ding{51} & black-box \\ \midrule
       CEMA         & Char+Word       & \ding{55}   & \ding{55} & \ding{51} & black-box \\ \bottomrule
       \end{tabular}%
       }
   \end{minipage}
\end{table*}

\begin{table*}[t]
\caption{Experiment results with improving the similarity threshold and adding attack methods for victim model A}\label{add_method}
\begin{tabular}{@{}cc|cccc@{}}
\toprule
Similarity Threshold & Method Number & ASR(\%)↑ & ASR(\%)↑ & BLEU↓ & Similarity↑ \\ \midrule
0.8                  & 3             & 73.57    & 62.27    & 0.14  & 0.93        \\
0.9                  & 6             & 67.16    & 58.89    & 0.16  & 0.95        \\
0.8                  & 3             & 80.80    & 65.40    & 0.15  & 0.93        \\
0.9                  & 6             & 78.25    & 61.15    & 0.16  & 0.94        \\ \bottomrule
\end{tabular}
\end{table*}

\section{Defense Method}\label{def}

We initiate an extensive exploration of defensive strategies to counter CEMA. In practical systems, we thoroughly investigate various defense mechanisms, including train-free adjustments(Preceding Language Modifier) and adversarial training.

\subsection{Preceding Language Modifier}\label{prece lang}

\begin{figure*}[h]
    \centering
    \includegraphics[width=0.93\textwidth]{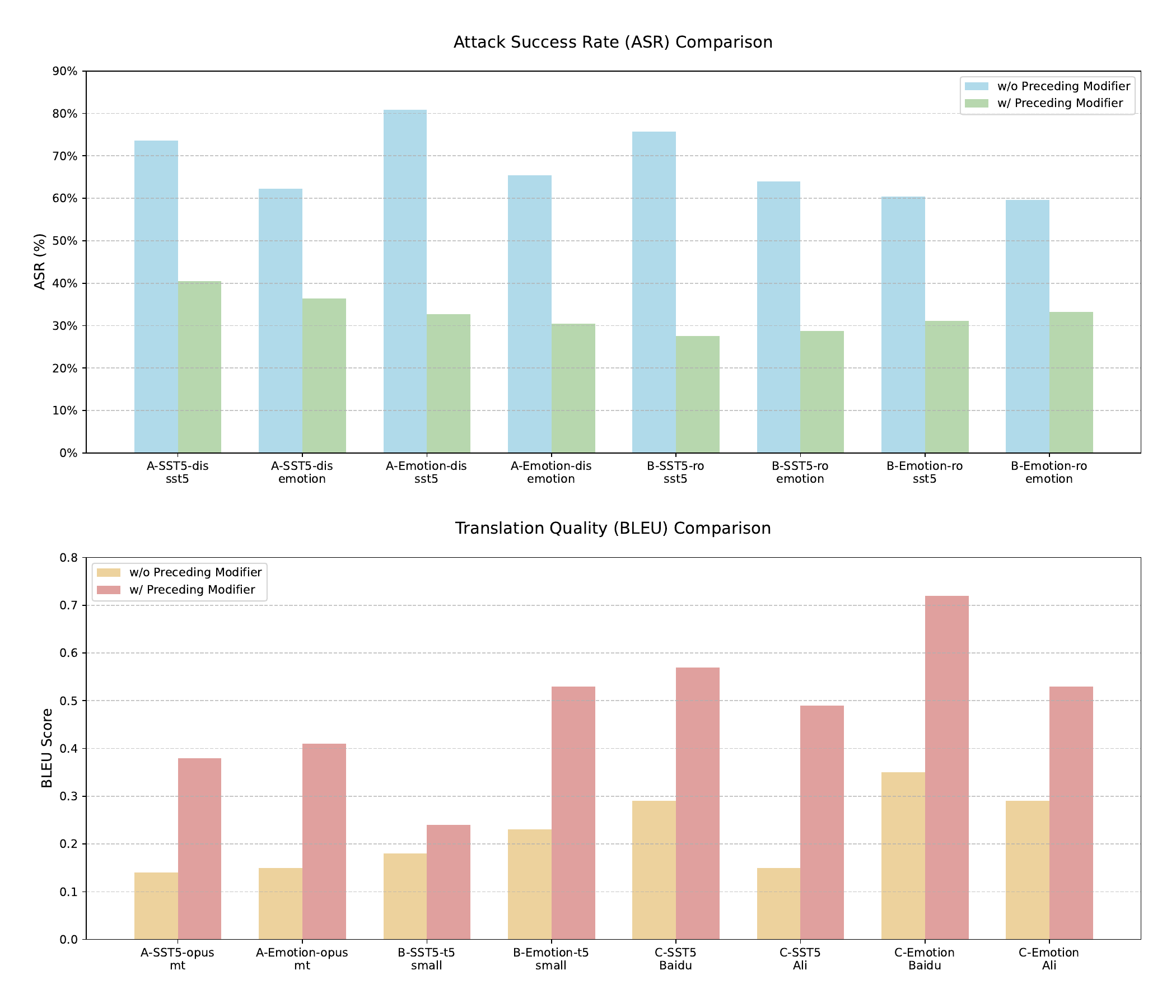} 
    \caption{The results of Preceding Language Modifier}\label{Modifier}
    
\end{figure*}

The victim models used in our study are after-trained models sourced from the Huggingface website, Ali Translator, and Baidu Translator. Since the training details of these pre-trained models are not publicly available, re-training them using adversarial training is infeasible. Consequently, we adopt training-free defense methods.  
Specifically, we implement the same approach proposed by ~~\cite{wang2023punctuation} and apply prompt learning techniques to large language models (LLMs) to mitigate adversarial text inputs. For this, we provide CoEdIT-XXL (a LLM used for correcting text errors). The prompt is as follows: ``Please revise the text for grammatical errors, improve the spelling, grammar, clarity, concision, and overall readability.'' The results are presented in Figure \ref{Modifier}.

``w/o'' indicates the absence of a defense method, whereas ``w'' denotes the use of the CoEdIT-XXL model as a modifier for defense.
Even after applying defense mechanisms using large language models, CEMA's attack effectiveness decreases but still maintains a significant level of performance.

\subsection{Adversarial Training}\label{adv tra}
\begin{figure}[h]
    \centering
    \includegraphics[width=0.50\textwidth]{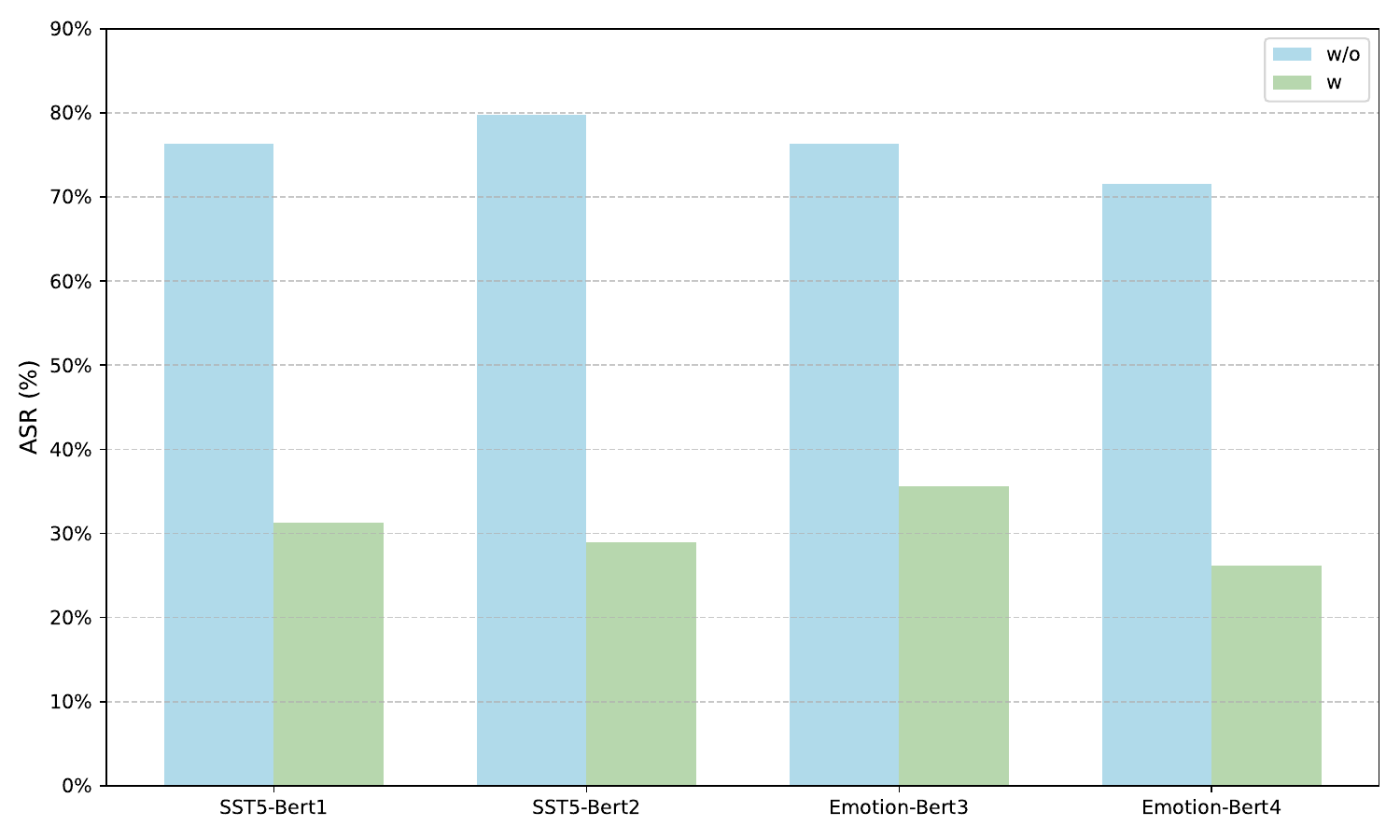} 
    \caption{The results of adversarial training.} 
    \label{AT} 
    
\end{figure}
We train four classification models as victim models and conduct adversarial training to evaluate the impact of adversarial training on CEMA's attack effectiveness. All four models are based on the BERT architecture and are labeled Bert1, Bert2, Bert3, and Bert4. Specifically, Bert1 and Bert3 are trained on the SST5 dataset, while Bert2 and Bert4 are trained on the Emotion dataset. The results are presented in Figure \ref{AT}. 
``w/o'' indicates the absence of adversarial training, while ``w'' represents the application of adversarial training. Although adversarial training reduces attack effectiveness, CEMA still demonstrates considerable performance.

\section{Definition of Text Classification Adversarial Examples and NMT Adversarial Examples}\label{def of text clas}

\subsection{Definition of NMT Adversarial Examples}

We define the source language space as $\mathcal{X}$ and the target language space as $\mathcal{Y}$, examining two NMT systems: the source-to-victim model $M_{x \to y}$, which maps $\mathcal{X}$ to $\mathcal{Y}$ to maximize $P(y_{\text{ref}} \mid x)$, and the target-to-source model $M_{y \to x}$, which performs the reverse mapping. After training, these models can reconstruct original sentences as $\hat{x} = g(f(x))$. We propose black-box adversarial testing for NMT using auxiliary data by selecting test sentences from $\mathcal{T} \subset \mathcal{X}$ and generating adversarial cases $\delta \in \Delta$ to perturb inputs $x' = x + \delta$ such that $f(x')$ diverges significantly from $f(x)$.

\textbf{NMT Adversarial Example:}
An NMT adversarial example is a sentence in 

\begin{equation}
\small
\hspace{-4.45mm}
    \begin{array}{c}
    \mathcal{A} = \left\{ x^{\prime} \in \mathcal{X} \mid \exists x \in \mathcal{T}\right\}, \\
    here \left\| x^{\prime} - x \right\| < \epsilon \wedge S_t\left(y, y_{\text{ref}}\right) \geq \gamma \wedge S_t\left(y^{\prime}, y_{\text{ref}}\right) < \gamma^{\prime} 
    \end{array}
    \hspace{-3mm}
\end{equation}
where function $f$ represents the NMT model. The variables $x$ and $x'$ represent the original text and the adversarial test case, respectively, while $y$ and $y'$ stand for their respective translations. In detail, $y = f(x)$ and $y' = f(x')$. The function $S_t(\cdot, \cdot)$ gauges the similarity between two sentences. Additionally, $\gamma$ and $\gamma'$ denote thresholds for acceptable translation quality. Translation quality is deemed unacceptable if $\gamma'$ drops below $\gamma$.

\subsection{Definition of Text Classification Adversarial Examples}
\textbf{Definition of Text Classification Adversarial Examples:}
Let \( X = \{x_1, x_2, \dots, x_n\} \) denote a set of text inputs, where each \( x_i \) is a text document (e.g., sentence or paragraph). Let \( f(\cdot) \) represent a text classification model, where:

\[
f: X \rightarrow Y
\]

is a mapping from the input space \( X \) to the label space \( Y \), with \( Y = \{y_1, y_2, \dots, y_m\} \) representing the set of possible class labels (e.g., positive, negative, neutral).

Given an input \( x \in X \) and its corresponding true label \( y_{\text{true}} = f(x) \), an \textit{adversarial example} \( \hat{x} \) is a perturbed version of the input \( x \) that is intentionally crafted to cause the model to misclassify it, while remaining perceptually and semantically similar to the original text. Formally, an adversarial example is defined as:

\[
\hat{x} = x + \delta
\]

where \( \delta \) is a small perturbation that satisfies:

\[
\|\delta\| \leq \epsilon
\]

Here, \( \|\delta\| \) represents the magnitude of the perturbation (e.g., measured in terms of the number of word substitutions or sentence modifications), and \( \epsilon \) is a threshold that bounds the maximum allowable perturbation.

Additionally, we impose a \textit{semantic similarity} constraint, ensuring that the perturbation \( \delta \) does not alter the meaning of the input significantly. This is formalized as:

\[
\text{Sim}(x, \hat{x}) \leq \gamma
\]

where \( \text{Sim}(x, \hat{x}) \) denotes a semantic similarity measure (such as cosine similarity) between the original input \( x \) and the adversarial example \( \hat{x} \), and \( \gamma \) is a predefined threshold that controls the acceptable level of semantic similarity. This ensures that the adversarial example \( \hat{x} \) remains semantically close to \( x \), while still leading to a misclassification.

The adversarial example \( \hat{x} \) causes the model to output a different class than the true label:

\[
f(\hat{x}) \neq y_{\text{true}} \quad \text{and} \quad f(x) = y_{\text{true}}
\]

Thus, the text classification adversarial example \( \hat{x} \) satisfies the following optimization problem:

\[
\hat{x} = \text{argmin}_{x' \in X} \mathcal{L}(f(x'), y_{\text{true}})
\]
\[
\text{subject to} \quad \|x' - x\| \leq \epsilon \quad \text{and} \quad \text{Sim}(x, x') \geq \gamma
\]

where \( \mathcal{L}(\cdot) \) is the loss function used to measure the discrepancy between the predicted label \( f(x') \) and the true label \( y_{\text{true}} \).

\section{The Experiment for Verifying Independence}\label{verify}

\begin{table*}[t]
\centering
\caption{The experimental results for verifying independence.}\label{verifying_independence}
\begin{tabular}{cc|ccccc}
\hline
Method A   & Method B   & P(A)    & P(B)    & P(A)*P(B) & P(AB)   & P(A)*P(B)-P(AB) \\ \hline \midrule
DWB        & FD         & 52.50\% & 40.50\% & 21.26\%   & 19.00\% & 2.26\%          \\
DWB        & Textbugger & 52.50\% & 72.50\% & 38.06\%   & 35.50\% & 2.56\%          \\
DWB        & Hotflip    & 52.50\% & 72.50\% & 38.06\%   & 37.00\% & 1.06\%          \\
DWB        & PSO        & 52.50\% & 76.50\% & 40.16\%   & 35.00\% & 5.16\%          \\
FD         & Textbugger & 40.50\% & 72.50\% & 29.36\%   & 31.00\% & -1.64\%         \\
FD         & Hotflip    & 40.50\% & 72.50\% & 29.36\%   & 31.50\% & -2.14\%         \\
FD         & PSO        & 40.50\% & 76.50\% & 30.98\%   & 30.50\% & 0.48\%          \\
Textbugger & Hotflip   & 72.50\% & 72.50\% & 52.56\%   & 57.50\% & -4.94\%         \\
Textbugger & PSO        & 72.50\% & 76.50\% & 55.46\%   & 54.00\% & 1.46\%          \\
Hotflip    & PSO        & 72.50\% & 76.50\% & 55.46\%   & 58.00\% & -2.54\%         \\ \hline
Average    &                            &                          &                          & 39.07\%   & 38.90\% & 0.17\%          \\ \hline
\end{tabular}
\end{table*}
We employ the DWB, FD, TextBugger, Hotflip, and PSO methods to generate adversarial examples in the subset of the test dataset. Since the exact success probabilities of each method's attacks are unavailable, we estimate these probabilities based on the observed frequency of successful attacks. In the table, we report the frequency \( P(AB) \) of both methods successfully attacking, as well as the individual success frequencies \( P(A) \) for Method A and \( P(B) \) for Method B. Our findings indicate that \( P(AB) \) closely approximates \( P(A) \times P(B) \), with the average deviation \( P(A) \times P(B) - P(AB) \) being just 0.17\%. The detailed experimental results are provided in Table \ref{verifying_independence}. Event independence is defined as the occurrence of event A having no effect on the occurrence of event B. Therefore, we assume that the success of adversarial examples generated by Method A does not influence the success of those generated by Method B.

\section{The Proof of Theorem \ref{max_prob2}}\label{proof_max_prob2}

We aim to prove that the probability of at least one successful attack, denoted as \( P(A^u) \), increases as the number of attack methods \( u \) increases. Specifically, we want to show that:

\begin{equation}
P(A^u) \leq P(A^{u+1}), \quad \text{for all} \quad u \geq 1,
\end{equation}

where \( A^u \) represents the event that at least one of the \( u \) attack methods succeeds, and \( A^{u+1} \) represents the event that at least one of the \( u+1 \) attack methods succeeds.

\textbf{Definitions}

1. Success Probability of Attack Methods:
   - Let \( p_i^j \) be the success probability of the \( j \)-th attack method of the victim text $x_i^*$.

2. Event \( A^u \):
   - \( A^u \) denotes the event that at least one of the \( u \) attack methods succeeds, i.e.,

   \begin{equation}
   A^u = A_1 \cup A_2 \cup \dots \cup A_u,
   \end{equation}

   where \( A_i \) is the event that the \( i \)-th attack method succeeds.

3. Event \( \overline{A^u} \):
   - \( \overline{A^u} \) denotes the event where none of the \( u \) attack methods succeeds, i.e.,

   \begin{equation}
   \overline{A^u} = \overline{A_1} \cap \overline{A_2} \cap \dots \cap \overline{A_u}.
   \end{equation}

   Therefore, the probability of at least one success is:

   \begin{equation}
   P(A^u) = 1 - P(\overline{A^u}).
   \end{equation}

4. Event \( A^{u+1} \):
   - Similarly, for \( u+1 \) attack methods, \( A^{u+1} \) is defined as:

   \begin{equation}
   A^{u+1} = A_1 \cup A_2 \cup \dots \cup A_{u+1},
   \end{equation}

   and \( \overline{A^{u+1}} \) represents the event where none of the \( u+1 \) attack methods succeeds:

   \begin{equation}
   \overline{A^{u+1}} = \overline{A_1} \cap \overline{A_2} \cap \dots \cap \overline{A_{u+1}}.
   \end{equation}

   Therefore, the probability of at least one success with \( u+1 \) attack methods is:

   \begin{equation}
   P(A^{u+1}) = 1 - P(\overline{A^{u+1}}).
   \end{equation}

\textbf{Goal}

We want to prove that

\begin{equation}
P(A^u) \leq P(A^{u+1}),
\end{equation}

i.e., the probability of at least one success increases as the number of attack methods \( u \) increases.

\textbf{Proof Process}

Step 1: Probability of Failure with \( u \) Attack Methods

The probability of failure with \( u \) attack methods is given by:

\begin{equation}
P(\overline{A^u}) = P(\overline{A_1} \cap \overline{A_2} \cap \dots \cap \overline{A_u}).
\end{equation}

- Independent Events: If the attack methods are independent, the failure probability is:

\begin{equation}
P(\overline{A^u}) = \prod_{j=1}^{u} (1 - p_i^j).
\end{equation}

- Dependent Events: If the attack methods are not independent, we have:

\begin{equation}
P(\overline{A^u}) \leq \prod_{j=1}^{u} (1 - p_i^j).
\end{equation}

Step 2: Probability of Failure with \( u+1 \) Attack Methods

The probability of failure with \( u+1 \) attack methods is:

\begin{equation}
P(\overline{A^{u+1}}) = P(\overline{A_1} \cap \overline{A_2} \cap \dots \cap \overline{A_{u+1}}).
\end{equation}

- Independent Events: If the attack methods are independent, the failure probability is:

\begin{equation}
P(\overline{A^{u+1}}) = \prod_{j=1}^{u+1} (1 - p_i^j).
\end{equation}

- Dependent Events: If the attack methods are dependent, we have:

\begin{equation}
P(\overline{A^{u+1}}) \leq \prod_{j=1}^{u+1} (1 - p_i^j).
\end{equation}

Step 3: Comparing \( P(A^u) \) and \( P(A^{u+1}) \)

We know:

\begin{equation}
P(A^u) = 1 - P(\overline{A^u}),
\end{equation}
\begin{equation}
P(A^{u+1}) = 1 - P(\overline{A^{u+1}}).
\end{equation}

- Independent Events:

\begin{equation}
P(A^u) = 1 - \prod_{j=1}^{u} (1 - p_i^j),
\end{equation}
\begin{equation}
P(A^{u+1}) = 1 - \prod_{j=1}^{u+1} (1 - p_i^j).
\end{equation}

Since \( \prod_{j=1}^{u+1} (1 - p_i^j) \leq \prod_{j=1}^{u} (1 - p_i^j) \), we have:

\begin{equation}
P(A^{u+1}) \geq P(A^u).
\end{equation}

- Dependent Events:

Similarly, since \( P(\overline{A^{u+1}}) \leq \prod_{j=1}^{u+1} (1 - p_i^j) \) and \( P(\overline{A^u}) \leq \prod_{j=1}^{u} (1 - p_i^j) \), we conclude that:

\begin{equation}
P(A^{u+1}) \geq P(A^u).
\end{equation}

\textbf{Conclusion}

Thus, we have proven that:

\begin{equation}
P(A^u) \leq P(A^{u+1}),
\end{equation}

i.e., the probability of at least one success increases as the number of attack methods \( u \) increases.

\textbf{Condition for Equality}

Equality \( P(A^u) = P(A^{u+1}) \) holds under the following conditions:

- The attack methods are completely independent and have identical success probabilities \( p_i^j = p_{u+1}^s \) for all \( i \), in which case adding an attack method does not change the overall failure probability.
- The newly added attack method does not contribute any additional success probability. This occurs when the new attack method is redundant or has no distinct contribution to the attack success.

\textbf{Discussion of Small Probability Events}

When the success probability of each attack method \( p_i^j \) is small (i.e., the attack methods have low success rates), in most cases, adding more attack methods leads to a significant increase in the probability of at least one success. This is because each additional attack method provides another opportunity for success, which, even with low individual probabilities, results in a cumulative increase in the overall success probability.

Therefore, in practical scenarios where each attack method has a small success probability, adding more methods increases the likelihood of at least one successful attack. This cumulative effect ensures that \( P(A^u) \) grows as \( u \) increases.

\textbf{Remark} { In Section \ref{Non-Independent} of the appendix, we examine the case of non-independence. We find that, in the non-independent scenario, using more methods to generate adversarial examples increases the likelihood of successfully attacking the victim model.}

Similarly, we can also prove
$P(B^u) \leq P(B^{u+1})$
\section{Supplementary Explanation for the Non-Independent Case in Section Candidate Adversarial Example Generation}\label{Non-Independent}

The probability of successfully attacking the victim model using adversarial examples generated by methods 1, 2, ..., \( n \) is greater than or equal to the probability of successfully attacking the victim model using adversarial examples generated by method 1 alone. This is because, when only method 1 is used, there is only one candidate adversarial example per victim text. In contrast, when \( n \) methods are employed, there are \( n \) candidate adversarial examples for each victim text, including the one generated by method 1. Therefore, the probability of successfully attacking the victim model using adversarial examples generated by \( n \) methods is greater than or equal to the probability of successfully attacking the victim model using adversarial examples from method 1 alone. The probabilities are equal only when method 1 achieves the maximum success rate for all victim texts. However, the SST5 and Emotion datasets contain 2,210 and 2,000 victim texts, respectively, making it unlikely that method 1 will achieve the maximum success rate across all victim texts. Thus, we conclude that, in most cases, the probability of successfully attacking the victim model using adversarial examples generated by \( n \) methods is greater than when using adversarial examples generated by method 1 alone.

Furthermore, based on this property, we can deduce that, in most cases, the probability of successfully attacking the victim model using adversarial examples generated by methods 1, 2, ..., \( n \) is greater than when using adversarial examples generated by methods 1, 2, ..., \( m \), where \( n > m \). In other words, employing more methods to generate adversarial examples increases the likelihood of a successful attack on the victim model.

\section{Multi-model Multi-task Learning (M3TL)}\label{M3TL}

Multi-model Multi-task Learning (M3TL) is a machine learning method that combines multiple learning models with multiple tasks. It is a combination of Multi-task Learning (MTL) and Multi-model Learning, aiming to improve model performance by jointly optimizing multiple tasks, especially when dealing with multiple related tasks.

\subsection{Key Concepts}

\subsubsection{Multi-task Learning (MTL)}
In traditional machine learning, each model typically handles a single task. In contrast, Multi-task Learning (MTL) involves jointly training multiple related tasks with a shared model. The goal is to allow the model to simultaneously optimize multiple objectives by sharing representations, knowledge, or parameters. Common applications include sentiment analysis and text classification, where the same features can be used for multiple tasks (e.g., predicting sentiment labels and classifying news articles). For instance, training a neural network to simultaneously perform two tasks: image classification and object detection.

\subsubsection{Multi-model Learning}
Unlike traditional single-model approaches, Multi-model Learning uses multiple independent or combined models to solve a problem. Each model may focus on different aspects of the problem or apply different algorithms to address the same task. For example, using multiple models such as neural networks, decision trees, and support vector machines to handle the same task, thereby leveraging the strengths of each model.

\subsubsection{Multi-model Multi-task Learning (M3TL)}
M3TL is a method that combines Multi-task Learning and Multi-model Learning. The core idea is to use multiple models (e.g., neural networks, decision trees, support vector machines, etc.) to learn multiple related tasks, with these models sharing some information or parameters. This means that during training, M3TL models handle multiple tasks and models simultaneously, enabling each model to learn across multiple tasks while sharing representations and knowledge between tasks.

\section{Low-quality Auxiliary Data}\label{Low-quality}
Initially, we treated the victim texts as auxiliary data. Now, we assume that the attacker cannot access the victim texts directly but can gather victim texts' relevant information 
and collect relevant  internet data as the auxiliary data. For example, the SST5 and Emotion datasets, both related to sentiment analysis but differing in label space and distribution. We use 100 unlabeled texts from the Emotion validation set as auxiliary data for the SST5 attack, and vice versa. The results in Table \ref{data-data}  show that despite limited auxiliary data and distribution differences, CEMA achieved a 66.45\% attack success rate and a BLEU score of 0.27. This demonstrates that an attacker only needs partial knowledge of the victim's data and can collect relevant information from the web to successfully attack the multi-task system using CEMA.
\begin{table*}[t] 
\centering
\caption{Attack performance of Low-quality Auxiliary Data. 
}\label{data-data}
\label{tab:my-table}
\resizebox{1\textwidth}{!}{%
\begin{tabular}{@{}cc|ccc|ccc|cc@{}}
\toprule
                                                       &                                                           & \multicolumn{3}{c|}{Victim Model A}                                                & \multicolumn{3}{c|}{Victim Model B}                                               & \multicolumn{2}{c}{Victim Model C}                                                                                                    \\ \cmidrule(l){3-10} 
                                                       &                                                           & dis-sst5 & dis-emotion & \begin{tabular}[c]{@{}c@{}}opus-mt\\ (en-zh) (A)\end{tabular} & ro-sst5 (A) & ro-emotion (A) & \begin{tabular}[c]{@{}c@{}}t5-small\\ (en-fr)\end{tabular} & \begin{tabular}[c]{@{}c@{}}Baidu Translate \\ (en-fr)\end{tabular} & \begin{tabular}[c]{@{}c@{}}Ali Translate \\ (en-zh)\end{tabular} \\ \midrule
\begin{tabular}[c]{@{}c@{}}Victim \\ Data\end{tabular} & \begin{tabular}[c]{@{}c@{}}Access \\ Data\end{tabular} &          ASR(\%)↑&             ASR(\%)↑&                                                           BLEU↓&         ASR(\%)↑&            ASR(\%)↑&                                                            BLEU↓& BLEU↓&                                                                  BLEU↓\\ \midrule \midrule
SST5                                                   & SST5                                                      & 73.6                 & 62.3                    & 0.14             & 75.7                & 64.0                   & 0.18              & 0.29                     & 0.15                                                                                                                     \\
                                                       & Emotion                                                   & 64.0                 & 60.8                    & 0.18             & 59.2                & 52.0                   & 0.22              & 0.36                     & 0.27                                                          \\ \midrule
\multirow{2}{*}{Emotion}                               & Emotion                                                   & 80.8                 & 65.4                    & 0.15             & 60.4                & 59.6                   & 0.23              & 0.35                     & 0.29                                         \\
                                                       & SST5                                                      & 66.4                 & 36.0                    & 0.21             & 48.8                & 46.4                   & 0.36              & 0.44                     & 0.42                                                            \\ \bottomrule
\end{tabular}
}

\end{table*}

\begin{table*}[t] 
\centering
\caption{The results of six tasks}\label{6-tasks}
\begin{tabular}{c|cccc|cc}
\hline
        & \multicolumn{4}{c|}{ASR(\%)}                  & \multicolumn{2}{c}{BLEU} \\ \hline
Data    & dis-emotion & ro-emotion & dis-sst5 & ro-sst5 & opus-mt    & t5-small    \\ \hline \midrule
SST5    & 75.91       & 74.90      & 67.04    & 62.82   & 0.18       & 0.22        \\
Emotion & 83.25       & 66.85      & 71.35    & 68.40   & 0.17       & 0.27        \\ \hline
\end{tabular}%
\end{table*}

\begin{table*}[t]
\centering
\caption{The results of translation, summary, and text-to-image tasks.}\label{t2i}
\resizebox{0.43\textwidth}{!}{
\begin{tabular}{@{}c|ccc@{}}
\toprule
Data                     & Task          & Metric                & Score \\ \midrule \midrule
\multirow{3}{*}{Pokemon} & Translation   & BLEU                  & 0.29  \\
                         & Summary       & RDP & 47\%  \\
                         & Text to Image & CDP  & 56\%  \\ \bottomrule
\end{tabular}
}
\end{table*}

\begin{table*}[t!]
\centering
\centering
\caption{
Performance of CEMA under different number setting of candidate adversarial examples.}\label{Example-Number}

\resizebox{1\textwidth}{!}{%
\begin{tabular}{@{}cccccccc@{}}
\toprule
\multirow{3}{*}{Data} & \multirow{3}{*}{\begin{tabular}[c]{@{}c@{}}Example\\ Number\end{tabular}} & \multicolumn{3}{c}{Victim Model A} & \multicolumn{3}{c}{Victim Model B} \\ \cmidrule(l){3-8} 
                      &                                                                           &            dis-sst5 (A) &           dis-emoton (A) &           opumt(En-Zh) (A) &            ro-sst5 (B)&           ro-emotion (B) &           t5-small(En-Fr) (B) 
\\ \cmidrule(l){3-8} 
                      &                                                                           &            ASR(\%)↑&           ASR(\%)↑&           BLEU↓        &            ASR(\%)↑&           ASR(\%)↑&           BLEU↓           
\\ \midrule \midrule
SST5                  
& \multicolumn{1}{c|}{3                               
}                                                                 & 73.6                 & 62.3                    & 0.14                 & 75.7                & 64.0                   & 0.18\\

& \multicolumn{1}{c|}{1                               
}                                                    & 50.4                 & 29.2                    & 0.30                  & 43.8                & 24.7                   & 0.35            
\\ \midrule
Emotion               
& \multicolumn{1}{c|}{3                               
}                                                     & 80.8                 & 65.4                    & 0.15                 & 60.4                & 59.6                   & 0.23\\
& \multicolumn{1}{c|}{1                               }                                                     & 29.2                 & 34.8                    & 0.31                 & 39.2                & 47.2                   & 0.39            \\ \bottomrule
\end{tabular}

}

\end{table*}

\begin{table*}[t] 
\centering
\centering
\caption{The results of few-shot and many-shot queries}\label{few-show-results}

\resizebox{1\textwidth}{!}{%
\begin{tabular}{@{}c|cccccc|cccccc@{}}
\toprule
\multirow{2}{*}{Model} & \multicolumn{6}{c|}{Victim Model A}                                                                                                                                                                                                                                                                              & \multicolumn{6}{c}{Victim Model B}                                                                                                                                                                                                                                                                             \\ \cmidrule(l){2-13} 
                       & dis-sst5 & dis-emotion & \multicolumn{1}{c|}{opus-mt} & dis-sst5 & dis-emotion & opus-mt & ro-sst5 & ro-emotion & \multicolumn{1}{c|}{t5-small} & ro-sst5 & ro-emotion & t5-small \\ \midrule
Data                   & \multicolumn{3}{c|}{SST5}                                                                                                                                         & \multicolumn{3}{c|}{Emotion}                                                                                                                 & \multicolumn{3}{c|}{SST5}                                                                                                                                        & \multicolumn{3}{c}{Emotion}                                                                                                                 \\ \midrule
Shot-Size              & ASR(\%)↑                                     & ASR(\%)↑                                        & \multicolumn{1}{c|}{BLEU↓}                                       & ASR(\%)↑                                     & ASR(\%)↑                                        & BLEU↓                                       & ASR(\%)↑                                    & ASR(\%)↑                                       & \multicolumn{1}{c|}{BLEU↓}                                        & ASR(\%)↑                                    & ASR(\%)↑                                       & BLEU↓                                        \\ \midrule \midrule
2000                   & 87.56    & 83.27       & \multicolumn{1}{c|}{0.1}     & 91.7     & 81.45       & 0.1     & 86.46   & 78.47      & \multicolumn{1}{c|}{0.15}     & 67.45   & 69.55      & 0.16     \\
1000                   & 83.04    & 76.76       & \multicolumn{1}{c|}{0.11}    & 88.25    & 76.15       & 0.12    & 84.16   & 73.49      & \multicolumn{1}{c|}{0.16}     & 66.35   & 67.05      & 0.17     \\
\textbf{100}                    & 73.57    & 62.27       & \multicolumn{1}{c|}{0.14}    & 80.8     & 65.4        & 0.15    & 75.66   & 64.01      & \multicolumn{1}{c|}{0.18}     & 60.4    & 59.60       & 0.23     \\
50                     & 63.71    & 45.64       & \multicolumn{1}{c|}{0.15}    & 71.05    & 53.55       & 0.18    & 71.69   & 59.38      & \multicolumn{1}{c|}{0.19}     & 58.65   & 57.9       & 0.24     \\
10                     & 38.38    & 32.06       & \multicolumn{1}{c|}{0.19}    & 43.35    & 37.7        & 0.21    & 59.28   & 46.51      & \multicolumn{1}{c|}{0.21}     & 46.15   & 41.75      & 0.27     \\ \bottomrule
\end{tabular}
}

\end{table*}

\end{document}